\documentclass[a4paper,10pt]{article}
\usepackage[utf8]{inputenc}
\usepackage[T1]{fontenc}
\usepackage{authblk}
\usepackage[english]{babel}
\usepackage{natbib}
\usepackage{graphicx}        
\usepackage{tabularx}
\usepackage{hyperref}
\usepackage{amssymb}
\DeclareGraphicsExtensions{.pdf,.png,.jpg,-eps-converted-to.pdf}

\chardef\us=`\_

\title{Analysis of large deflections of prominence-CME events during the rising phase of solar cycle 24}

\author[1]{M.V.~Sieyra}
\author[2]{M.~C\'ecere}
\author[1]{H.~Cremades}
\author[1]{F.A.~Iglesias}
\author[2]{A.~Sahade}
\author[3,4]{M.~Mierla}
\author[5]{G.~Stenborg}
\author[2]{A.~Costa}
\author[3]{M.~West}
\author[3]{E.~D'Huys}
\affil[1]{Universidad Tecnol\'ogica Nacional\,--\,Facultad Regional Mendoza, CONICET, CEDS, Rodriguez 243, Mendoza, Argentina}
\affil[2]{Instituto de Astronom\'ia Te\'orica y Experimental, CONICET-UNC,  C\'ordoba, Argentina}
\affil[3]{Solar-Terrestrial Center of Excellence - SIDC, Royal Observatory of Belgium, Brussels, Belgium}
\affil[4]{Institute of Geodynamics of the Romanian Academy, Bucharest, Romania}
\affil[5]{Space Science Division, U.S. Naval Research Laboratory, Washington, DC 20375, USA}

\begin{document}

\maketitle

\begin{abstract}
Motivated by the need to improve the ability to forecast whether a certain coronal mass ejection (CME) is to impact Earth, and by the insufficiency of statistical studies that analyze the whole erupting system with the focus on the governing conditions under CME deflections, we performed a careful analysis of 13 events along a one-year time interval showing large deflections from their source region. We used telescopes imaging the solar corona at different heights and wavelengths on board the \textit{Project for Onboard Autonomy 2} (PROBA2), \textit{Solar Dynamics Observatory} (SDO), \textit{Solar TErrestrial RElations Observatory} (STEREO), \textit{Solar and Heliospheric Observatory} (SOHO) spacecraft and from \textit{National Solar Observatory} (NSO). By taking advantage of the quadrature position of these spacecraft from October 2010 until September 2011, we inspected the 3D trajectory of CMEs and their associated prominences with respect to their solar sources by means of a tie-pointing tool and a forward model. Considering the coronal magnetic fields as computed from a potential field source surface model, we investigate the roles of magnetic energy distribution and kinematic features in the non-radial propagation of both structures. The  magnetic environment present during the eruption is found to be crucial in determining the trajectory of CMEs, in agreement with previous reports.
\end{abstract}



\section{Introduction}
\label{s:introduction} 

Coronal mass ejections (CMEs) are large scale phenomena that constantly erupt from the solar surface traveling through the interplanetary space. They constitute one of the primary drivers of space weather events, such as geomagnetic storms, solar energetic particles, etc. When assessing the capacity of a particular CME to affect Earth or another natural or artificial object, it is, of course, important to have knowledge of its magnetic field orientation and other energy-related parameters. However, in the first place, it is of utmost importance to correctly ascertain its propagation direction and size, so as to determine whether the impact will take place at all and will enable us to perform more accurate space weather predictions. 

It is well-known that CMEs not always propagate radially outward from their source regions \citep[e.g.,][]{macqueen1986, gosling1987, vandas1996, cremades2004, gui2011, rollett2014, kay2015, mostl2015} and determining their direction of propagation may not be straightforward from a single viewpoint,  particularly if directed towards it. Since the launch of the \textit{Solar TErrestrial RElations Observatory} \citep[STEREO,][]{kaiser2008} together with the development of various reconstruction tools \citep[e.g.,][]{mierla2008, mierla2010, maloney2009, dekoning2009, temmer2009, srivastava2009, liewer2009, thernisien2009, thernisien2011jastp, liu2010}, it is possible to obtain three-dimensional (3D) information of CMEs and their associated prominences. This allows us to determine the deflection in latitude as well as in longitude from the source location for both structures. It has also provided new insights into the relationship between various features associated with filaments and CME eruptions.

Moreover, to date it has not been possible to predict before eruption whether a specific CME, to be born in a particular region on the Sun under specific environmental conditions, is to be deflected and to what extent. Although there are some studies in this direction \citep[e.g.,][]{kay2015, zhuang2017} the detailed analysis on the causes of deflection are focused only on case studies \citep[e.g.][]{gui2011, panasenco2013, liewer2015, kay2017, cecere2020}.

It has been shown that in activity-minimum years there is a systematic deflection to lower latitudes and no systematic trend  at times of high activity \citep[e.g.,][]{cremades2004, wang2011}. During solar minimum, the heliospheric current sheet (HCS) remains flat at low latitudes, so predominantly latitudinal deflections occur towards the equator. During other times of the solar cycle, the HCS transitions to a more complex configuration, which would allow deflections to have a more significant longitudinal component, as suggested by \citet{kay2015}.

It is also widely known that CMEs propagate non-radially away from nearby coronal holes and toward regions of low magnetic energy. For example \citet{cremades2006} found a good correspondence between the deflection of CMEs and the total area of coronal holes (CHs), suggesting that the neighboring CHs affect the evolution of CMEs near the Sun.  \citet{gopalswamy2009} also suggested that CMEs could be deflected by the associated CHs and claimed that the open flux from these structures acted as magnetic walls, constraining CME propagation. The work performed by \citet{shen2011} showed that the trajectory of the analyzed CMEs were influenced by the background magnetic field and that they are likely to deflect to the nearby region with lower magnetic energy density. \citet{gui2011}, extending the work of \citet{shen2011} to ten CMEs,  and analyzed the deflection in both latitude and longitude. Aside from verifying the previous results, they found a positive correlation between the deflection rate and the strength of the gradient of the magnetic energy density. 

In addition to these causes, recent studies have demonstrated that CMEs are also deflected by strong magnetic fields from active regions in the locations of CME source \citep[e.g.,][]{mostl2015, wang2015, kay2015, kay2017}, with the magnitude of the deflection being inversely related to CME speed and mass. This was previously suggested by \citet{xie2009} and \citet{kilpua2009}. Slower and wider CMEs deflect toward the equator during solar minimum while  faster and narrower CMEs deflect less, even in some cases they propagate radially from it source active region. It was suggested that slow and wider CMEs cannot penetrate through the background  overlying coronal fields, but are channeled toward the streamer belt. Also the background fast solar wind can inhibit the latitudinal expansion of the CME in the corona \citep[e.g.,][]{cremades2006} and interact with CMEs at large distances \citep[e.g.][]{isavnin2014} where the magnetic forces from the background are negligible. Recent numerical research by \citet{Zhuang2019} supports CME deflection in interplanetary space relative to the difference between CME and solar wind  speed, i.e. the greater the difference, the larger the deflection. Interactions between multiple CMEs/ICMEs can also cause deflections, mainly longitudinal \citep[e.g.,][]{lugaz2012, shen2012, liu2012, liu2014}. Summarizing, the rate and amount of CME deflection is believed to be controlled by the strength and distribution of the background magnetic field, and the mass, size, and speed of the CME relative to the solar wind. Hence, both the global and local configuration of the Sun's magnetic field together with intrinsic CME properties would have crucial importance on the degree and direction of deflection.

At the same time, prominence deflection and rolling motions during the process of eruption has received less attention, though there has been some work along this line \citep[e.g.][]{filippov2001, martin2003, panasenco2008, bemporad2009, panasenco2011, pevtsov2012, liewer2013}. The filament, the channel encompassing the polarity reversal boundary, the overlying arcade, and the CME itself are all part of one linked magnetic system \citep{martin2008, pevtsov2012}. The filament eruption and the CME are two manifestations of the same underlying magnetic phenomenon,  thus by studying filament eruptions we can better understand CME triggering and improve our ability to predict it. Very few studies combine the dynamics of the prominence and CME. For example \citet{panasenco2013} demonstrate that major twists and non-radial motions in erupting filaments and CMEs are typically related to the larger-scale ambient conditions around the eruptive events. They found that the non-radial propagation of both structures is correlated with the presence of nearby coronal holes and are guided towards weaker field regions, namely null points existing at different heights in the overlying magnetic configuration. The CME propagates in the direction of least resistance, which is always away from the coronal hole, and the non-radial direction of the erupting filament system is caused either by the open coronal hole magnetic field near the filament channel or by other strong magnetic field which might be in the neighborhood of the eruption. They also found that the non-radial motion of the prominence is greater than that of the CME. Another study that considers the magnetic background surrounding the source region, is reported by  \citet{liewer2015}. They analyze the coronal magnetic pressure forces acting on CMEs at different heights in the corona and also consider the non-radial propagation below the coronagraph field of view (FOV). They conclude that non-radial propagation can result not only from large-scale coronal fields, but also from initial asymmetric expansion caused by the nearby strong active-region fields. They found that CMEs propagate through the weak field region around the HCS and do not follow the shortest path to the HCS but the path depends on the local and global gradients in the magnetic pressure.

Given the importance of understanding non-radial propagation to improve our ability to forecast
whether or not a CME will impact Earth, and motivated by the lack of statistical studies that analyze the whole erupting system focusing on the main causes of deflection, we perform a systematic study of the deflection of CMEs and their associated prominences with respect to their solar sources. Taking advantage of the spacecraft fleet dedicated to study solar activity including the \textit{Solar and Heliospheric Observatory}  \citep[SOHO,][]{domingo1995}, the \textit{Solar Dynamics Observatory} \citep[SDO,][]{pesnell2012}, the  \textit{Solar-Terrestrial Relations Observatory} \citep[STEREO,][]{kaiser2008} and the \textit{National Solar Observatory} (NSO) together with the reconstruction methods mentioned above, we determine the trajectory of CMEs and their corresponding prominences. Considering the coronal magnetic fields as computed from a \textit{Potential Field Source Surface} model \citep[PFSS,][]{schrijver2003} we attempt to investigate the roles of magnetic energy distribution and kinematic features in the non-radial propagation of both structures.

The methodology, including the identification criteria used to compile the analyzed events, the methods to determine the trajectory of the prominences and CMEs and the estimation of the magnetic energy at different heights, is described in Section \ref{s:obsmeth}. The obtained results, both in relation to the kinematics and the magnetic environmental conditions  are presented in Section \ref{s:results} together with a detailed analysis of some specific cases in Section \ref{ss:anomalous_cases}. Finally, we discuss and summarize our main findings in Section \ref{s:discussion}.

\section{Observations and Methodology}
\label{s:obsmeth}

\subsection{Data and events' selection}
\label{ss:obs}

Since our main interest entails in the investigation of CMEs having large deflections with respect to their solar sources, we pre-select candidate events for the study by means of the following procedure. As a first step, we considered all filament eruptions reported by the  \href{http://aia.cfa.harvard.edu/filament/}{\textit{AIA Filament Eruption Catalog}} \citep{mccauley2015} from October 2010 until September 2011. We chose this time interval because the quadrature location between spacecraft on the Sun-Earth line and the STEREO twin probes provides a better three-dimensional perspective of the prominences and associated CMEs. Out of the 183 filament eruptions reported by the \textit{AIA Filament Eruption Catalog} during that time interval, we found 118 events that resulted in CMEs detected in the field of view of white-light coronagraphs. The erupting filament-CME associations where performed with the aid of the \href{https://cdaw.gsfc.nasa.gov/CME_list/}{\textit{SOHO/LASCO CME Catalog}} \citep{yashiro2004}. Next, to pre-select candidate events having large deflections, we checked for large differences ($\gtrsim 20^\circ$) between position angles of the filaments before erupting and of their ensuing CMEs, both angles measured on the plane of sky from the same viewpoint and counterclockwise from the Solar North. We chose a value of $20^\circ$ in agreement with the average unsigned deflection found by \citet{cremades2006}.

\begin{figure}
    \centerline{
    \includegraphics[width=\textwidth]{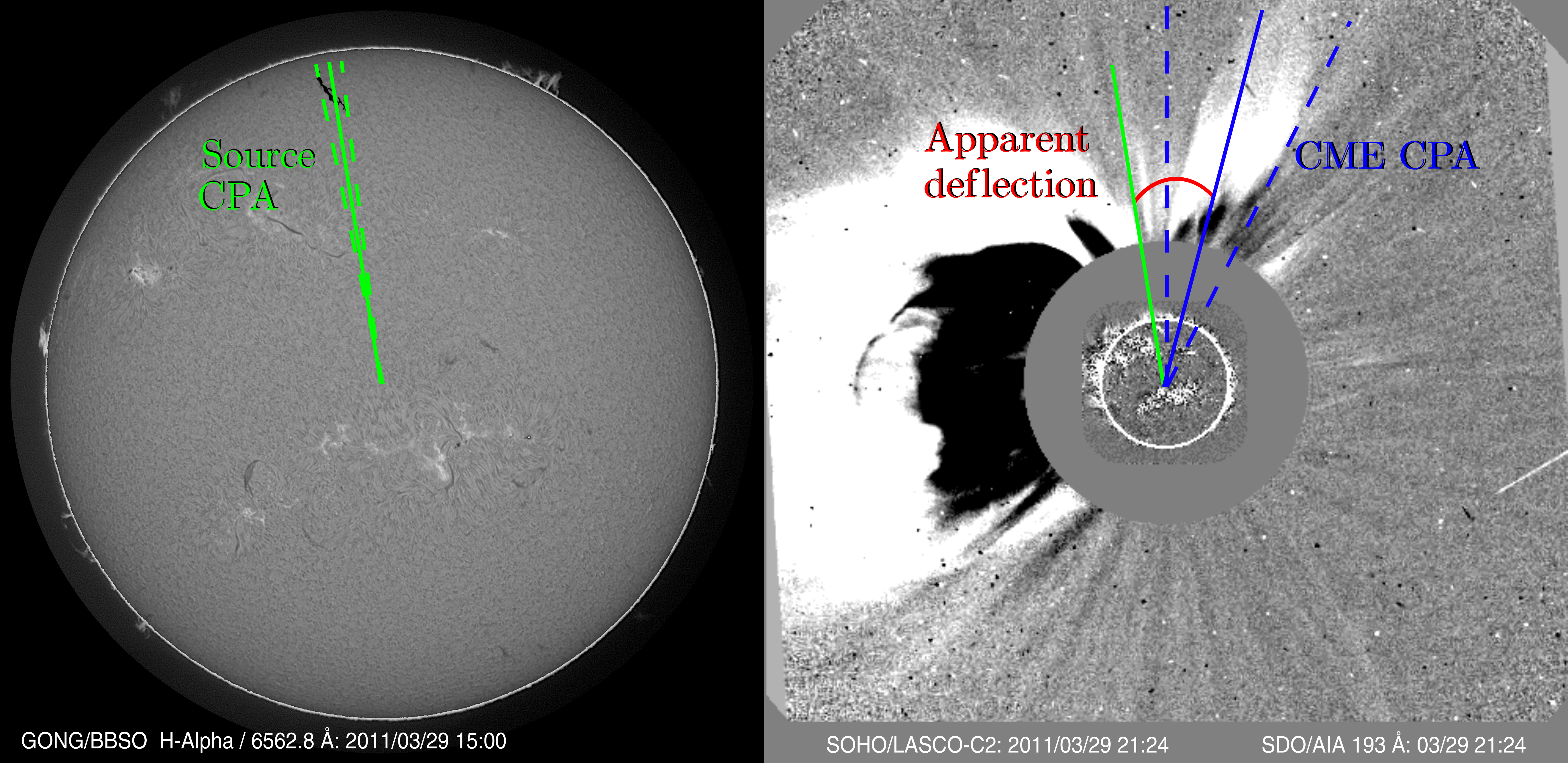}}
    \caption{Apparent deflection from Earth's view defined by the difference in position angle between the middle point of the source region (Source CPA , left panel) and the central position angle of the CME (CME CPA, right panel). The source is seen in $H_\alpha$ image from BBSO at 15:00 UT and the CME image is taken from SOHO/LASCO-C2 at 21:24 UT on 29 March 2011.}
    \label{fig:app_defl}
\end{figure}

\begin{figure}
    \centering
    \includegraphics[width=0.4\textwidth, angle=90]{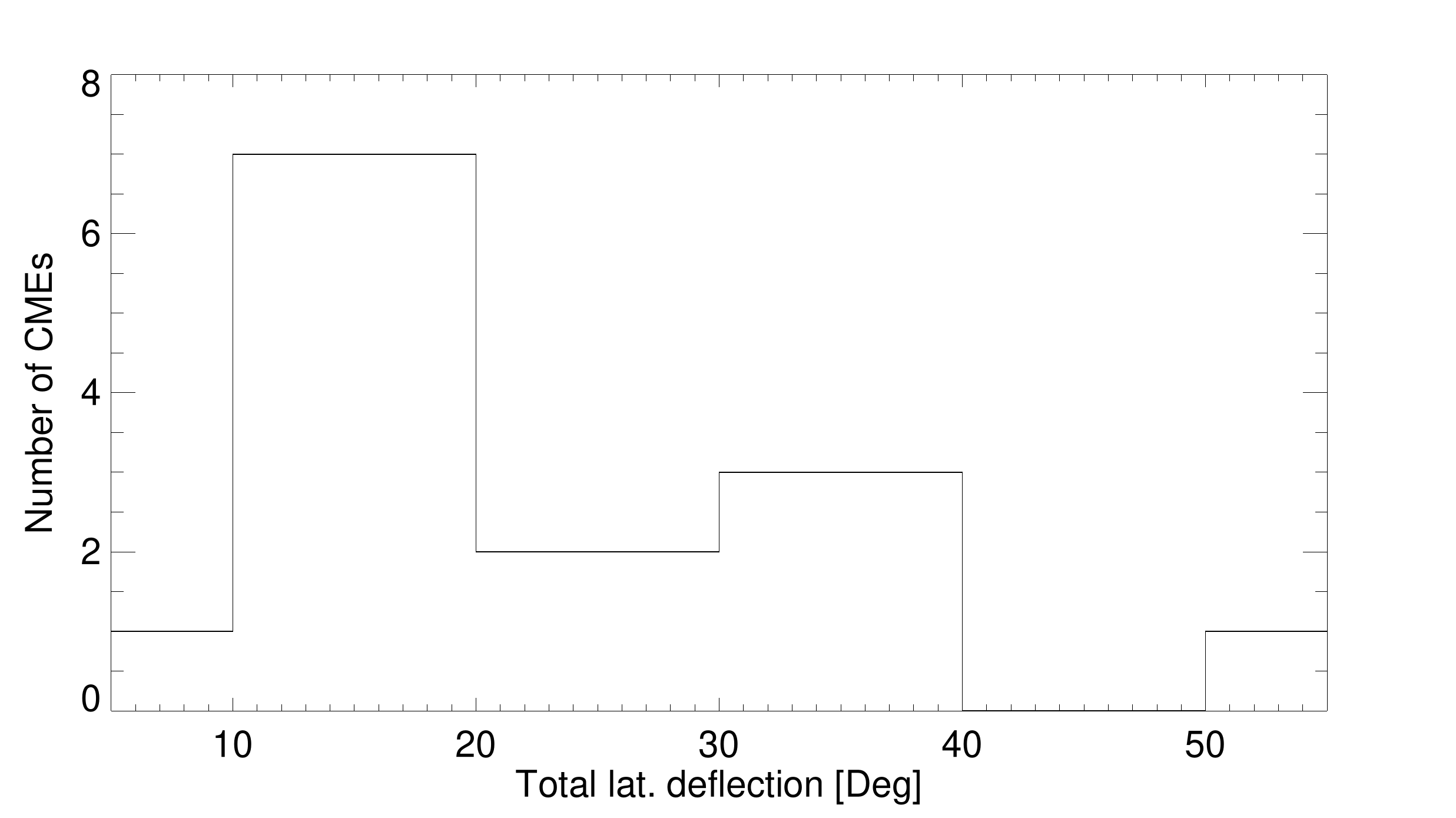}
    \includegraphics[width=0.4\textwidth, angle=90]{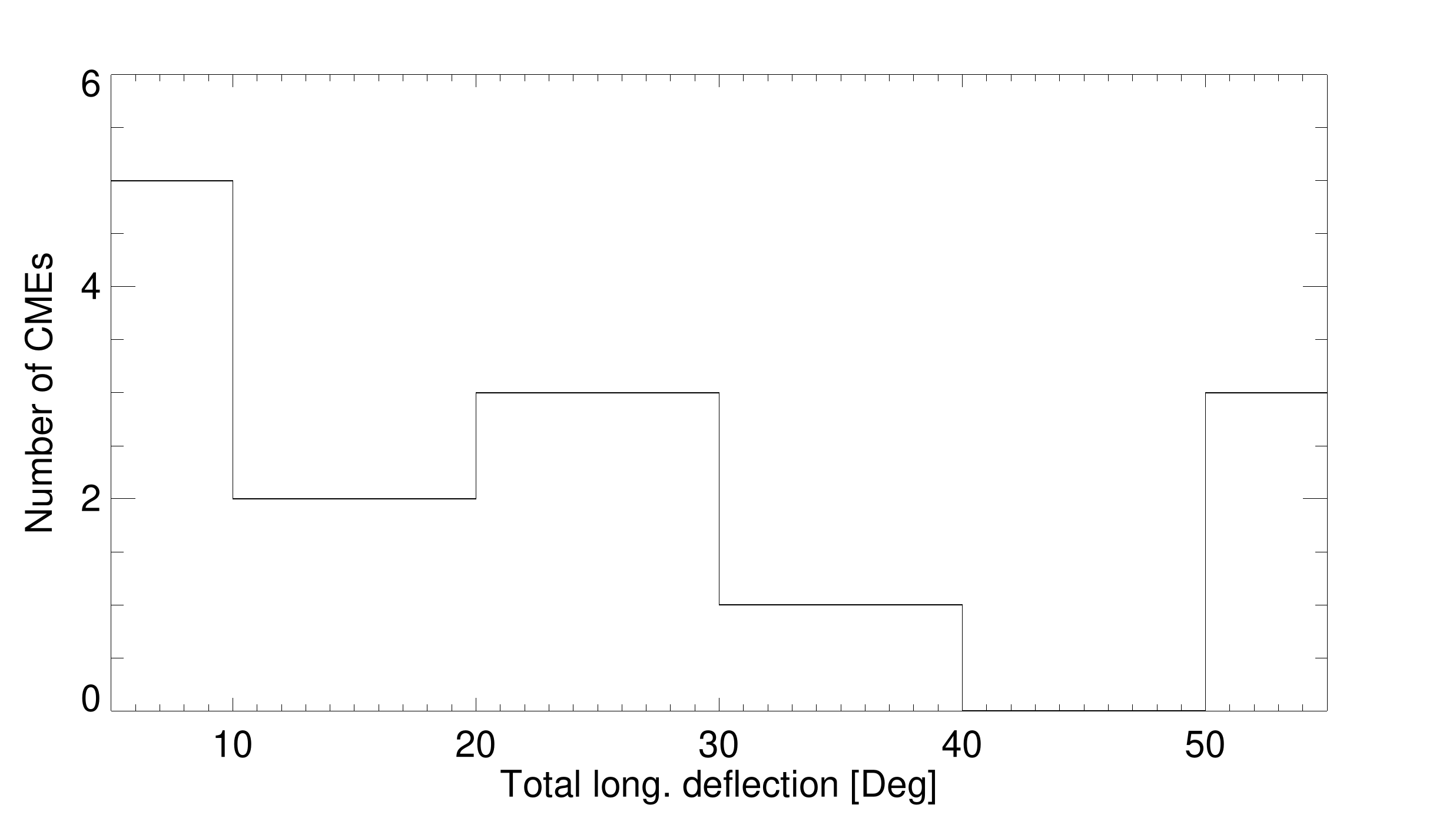}
    \includegraphics[width=0.4\textwidth, angle=90]{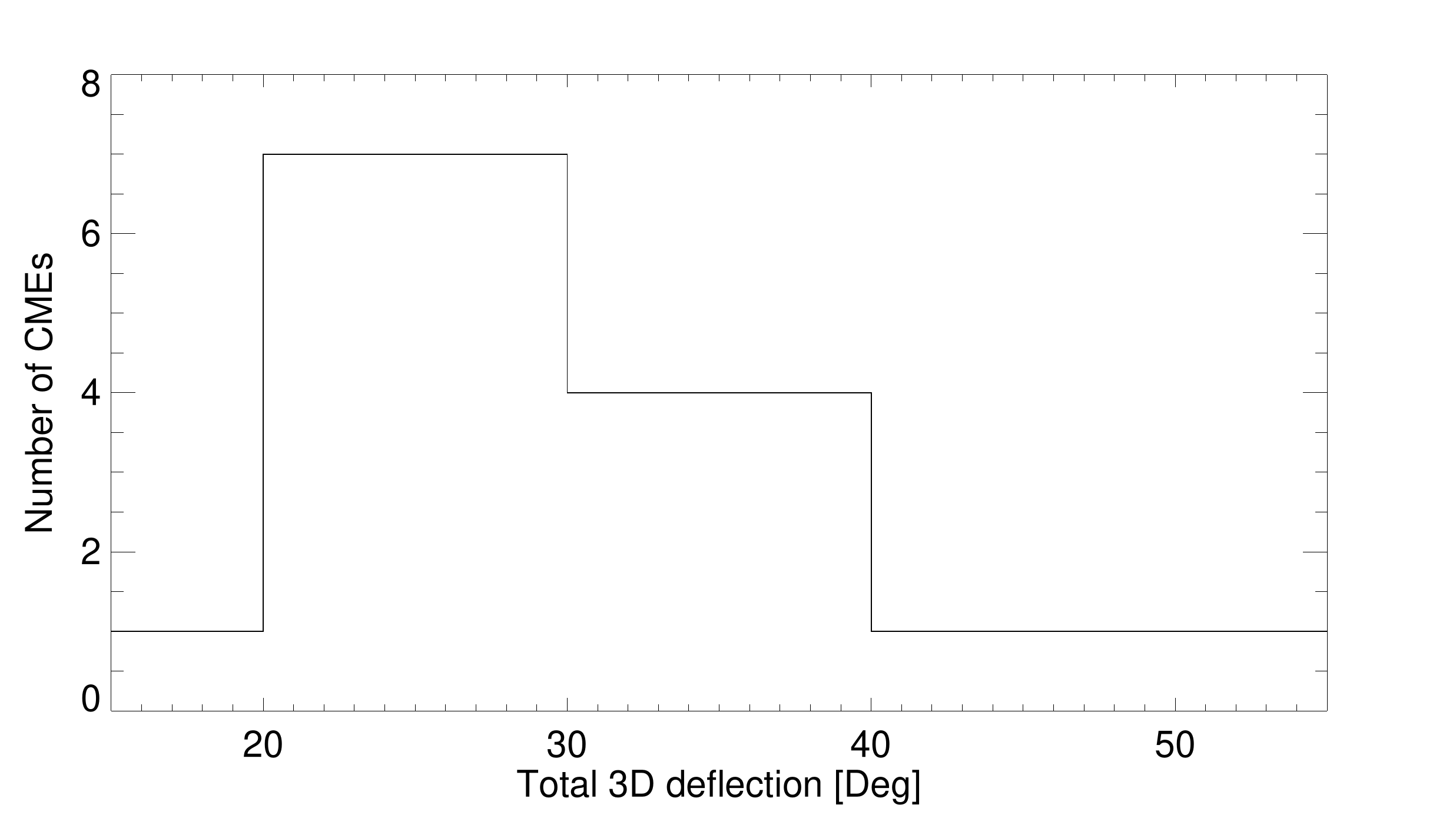}
    \caption{Distribution of the deflection in latitude (upper panel), longitude (central panel) and 3D (bottom panel) for the analyzed events. The deflections shown here were calculated considering the central coordinates of the source region and the coordinates of the associated CME apex at its highest measured point.}
    \label{fig:coord_hist}
\end{figure}

To measure the position angle of the central point of the filament (Source CPA) in its pre-eruptive phase we  used images in $H{\alpha}$ from the \textit{Global Oscillation Network Group} \citep[GONG,][]{kennedy1994} from the \textit{National Solar Observatory Integrated Synoptic Program} (NISP). The GONG network of instruments is hosted by observatories geographically distributed around the Earth: Big Bear Solar Observatory, California; Mauna Loa Solar Observatory, Hawaii; Learmonth Solar Observatory, Australia; Udaipur Solar Observatory, India; Observatorio del Roque de los Muchachos, Canary Islands, Spain; and Cerro Tololo Interamerican Observatory, Chile. Whenever the source pre-eruptive filament could not be fully detected in $H{\alpha}$, either because it was too faint in this wavelength or its location was not on the visible side as seen from Earth, we used images from the extreme-ultraviolet (EUV) telescopes, namely the \textit{Atmospheric Imaging Assemby} \citep[AIA,][]{lemen2012} onboard SDO, and the \textit{Sun-Earth Connection Coronal and Heliospheric Investigation EUV Imager} \citep[SECCHI-EUVI,][]{howard2008} onboard the STEREO twin spacecraft. CME central position angles (CPA) were measured on images from LASCO-C2 \citep[\textit{Large Angle and Spectrometric Coronagraph Experiment},][]{brueckner1995} and SECCHI-COR2 at a height of $\sim 5\,R_\odot$ (projected on the plane of sky), assuming that CMEs are fully developed and their evolution is self-similar at this height.
A scheme that clarifies the pre-selection criterion is presented in Figure~\ref{fig:app_defl}. The left panel of the figure displays an $H_\alpha$ image from \textit{Big Bear Solar Observatory} (BBSO), where the green dashed lines encompass the filament that erupts later, and the green solid line indicates the Source CPA considered as source of the CME. The right panel shows the associated CME as seen by SOHO/LASCO-C2, with the blue dashed lines encompassing the CME's angular width and the solid line its  CME CPA. The difference between these CPAs, shown in red, represents the deflection projected onto the plane of sky of the instrument (apparent deflection). 
After this pre-selection of events whose projected deflection on the basis of CPAs is greater than $20^\circ$, we further constrain our sample by examining whether that apparent, \textit{i.e.} projected, deflection corresponds to a real deflection. The overall ``real'' (\textit{i.e.} 3D) deflection is defined by the difference in latitude ($\Delta \Theta$) and Carrington longitude ($\Delta \Phi$) between the central coordinates of the source region, \textit{i.e.} those of the filament in its pre-eruptive state, and the coordinates of the resulting CME at the greatest measured height. The methods used to deduce the 3D coordinates (latitude, longitude and height) of CMEs and source regions, among another parameters, are described in Section \ref{ss:meth}. On the basis of spherical trigonometry, the 3D total deflection is thus defined as:   
\begin{equation}
\Delta \Psi=\arccos (\sin (\Theta_{sr})\sin (\Theta_{cme}) +\cos (\Theta_{sr})\cos (\Theta_{cme}) \cos(|\Phi_{sr}-\Phi_{cme}|)),
\label{eq:defl}
\end{equation}
where sub-indexes $sr$ and $cme$ indicate a coordinate that is associated to the source region and the CME, respectively. A given event is selected for further analysis only if $\Delta \Psi \gtrsim 20^\circ$. Out of the 118  events reported by the \textit{AIA Filament Eruption Catalog} during the investigated time interval, 23 were initially pre-selected as they exhibited a projected deflection $|\Delta \textrm{CPA}| \gtrsim 20^\circ$; but only 13 of these events yielded overall 3D deflections $\Delta \Psi \gtrsim 20^\circ$ according to our measurement method. The 10 remaining events were discarded due to several reasons: either their 3D deflections were small ($\Delta \Psi < 20^\circ$), or there were data gaps in COR2 or LASCO, or the CMEs were too faint to deduce their latitude and longitude applying the method described in the following section. The 13 selected events that satisfy  $\Delta \Psi \gtrsim 20^\circ$ are summarized in Table \ref{tab:events}. The table indicates CPAs and coordinates (latitude $\Theta$ and longitude $\Phi$) of the source region and CME, the difference between these measurements and the obtained overall 3D deflection. We also show the distribution of the resulting deflection in latitude, longitude and 3D for the selected events in Figure~\ref{fig:coord_hist}. Most of the events present latitudinal deflection between 10 and 20$^\circ$ and a longitudinal deflection lower than 10$^\circ$, while there are fewer events that exhibit deflections larger than 50$^\circ$ in both coordinates. The total 3D deflection $\Delta \Psi$ results mainly between 20$^\circ$ and 30$^\circ$. This figure also indicates that our sample of events presents latitudinal and longitudinal deflections in similar ranges.
\begin{table}
{\tiny
\centering
 \begin{tabularx}{\hsize}{ccccccccccccc}
 \hline \\ 
 \multicolumn{2}{c}{Catalog} & \multicolumn{3}{c}{SR} & \multicolumn{4}{c}{CME} & \multicolumn{4}{c}{Deflection} \\
 \hline \\
ID & Date & CPA & $\Theta$ & $\Phi$ & Time$^*$ & CPA & $\Theta$ & $\Phi$ & $\Delta \textrm{CPA}$ & $\Delta \Theta$ & $\Delta \Phi$ & $\Delta \Psi$ \\
(1)&(2)&(3)&(4)&(5)&(6)&(7)&(8)&(9)&(10)&(11)&(12)&(13)\\
 \hline \\
 118 & 2010-11-24 & 344$^b$ & 62  & 76  & 07:36 & 325 & 43   & 67   & -19 & -19 & -9   & 20  \\  
 132 & 2010-12-16 & 298$^d$ & 29  & 110 & 08:48 & 326 & 43   & 138  & 28  & 14  & 28   & 26  \\ 
 136 & 2010-12-23 & 212$^a$ & -53 & 66  & 05:00 & 234 & -17  & 33   & 22  & 36  & -33  & 44 \\
 142 & 2011-01-02 & 209$^b$ & -58 & 347 & 06:12 & 255 & -5   & 347  & 46  & 53  & 0    & 53  \\
 159 & 2011-01-30 & 320$^a$ & 25  & 250 & 18:36 & 278 & 7    & 272  & -42 & -18 & 22   & 28 \\
 180 & 2011-02-25 & 34$^a$  & 43  & 208 & 08:00 & 348 & 45   & 263  & 46  & 2   & 55   & 39 \\
 196 & 2011-03-27 & 354$^a$ & 68  & 205 & 20:12 & 324 & 51   & 255  & -30 & -17 & 50   & 29 \\
 197 & 2011-03-29 & 9$^a$   & 51  & 169 & 20:36 & 347 & 64   & 224  & 22  & 13  & 55   & 31 \\
 216 & 2011-05-13 & 216$^a$ & -38 & 357 & 18:48 & 254 & -8   & 351  & 38  & 30  & -6   & 30 \\
 251 & 2011-07-07 & 119$^a$ & -19 & 252 & 13:25 & 99  & 1    & 244  & -20 & 20  & -8   & 21 \\
 274 & 2011-08-10 & 310$^c$ & 41  & 43  & 05:00 & 334 & 64   & 49   & 24  & 23  & 6    & 23 \\
 276 & 2011-08-11 & 287$^b$ & 18  & 291 & 10:36 & 267 & -1   & 269  & 20 & -19  & -22  & 29 \\
 286 & 2011-09-08 & 60$^c$  & 28  & 226 & 06:12 & 38  & 47   & 240  & -22 & 19  & 14   & 22 \\
 \hline \\
 \multicolumn{13}{l}{$^*$ First LASCO-C2 appearance time [UT].} \\
 \multicolumn{13}{l}{$^a$ Measured using $H_\alpha$ images.} \\
 \multicolumn{13}{l}{$^b$ Measured using SDO/AIA images.} \\
 \multicolumn{13}{l}{$^c$ Measured using STEREO-A/EUVI.} \\
 \multicolumn{13}{l}{$^d$ Measured using STEREO-B/EUVI.} \\
 \end{tabularx}}
 \caption{The 13 selected events that satisfy $|\Delta \textrm{CPA}| \gtrsim 20^{\circ}$ and $\Delta \Psi \gtrsim 20^{\circ}$ between October 2010 and September 2011. The first two columns contain the AIA Filament Eruption Catalog ID and the date of the reported event, columns 3\,--\,5 indicate the source region location (CPA, latitude and Carrington longitude), columns 6\,--\,9 exhibit the CME first time appearance in LASCO-C2 and location parameters, while columns 10\,--\,13 show the resulting deflection in position angle, latitude, and longitude, as well as the total deflection.}
  \label{tab:events}
 \end{table}

\subsection{Determination of 3D coordinates and tracking}
\label{ss:meth}
\subsubsection{Coordinates}

After the pre-selection procedure we determined 3D coordinates of the source region and ensuing CME, to ascertain whether the apparent deflection was indeed related to a real deflection similar to or larger than $20^\circ$. To determine the 3D coordinates of the source region, which we defined as the central position coordinates of the filaments in their pre-eruptive state, we used $H_\alpha$ images from the \href{http://halpha.nso.edu/archive.html}{NSO/GONG \textit{H$\alpha$ Archive}} using \href{http://www.lmsal.com/solarsoft/}{\textit{SolarSoft}} standard procedures. In those cases where the filament was not clearly discernible in that wavelength or it was too close to the limb or back-sided, we measured the coordinates in SDO/AIA or STEREO/EUVI 304\,\AA{} images by means of the \href{https://www.jhelioviewer.org/}{\textit{JHelioviewer}} \citep{jhelioviewer} image visualization tool.

As central 3D coordinates of each CME, we considered those yielded by the Graduated Cylindrical Shell (GCS) forward model \citep[]{thernisien2006, thernisien2009} at the highest possible altitude, dependent on the particular visibility conditions of each case. This method reproduces the large-scale structure of a flux rope-like CME by modeling its outer envelope as a hollow croissant-like shape. Briefly, the model consists of a tubular section forming the main body of the structure attached to two cones that correspond to the ``legs'' of the CME. Fitting the GCS model to the CMEs in the SOHO/LASCO and STEREO/COR2 coronagraph images enables not only to estimate their 3D direction of propagation (longitude and latitude), but also their apex height, half angular width, tilt angle of the symmetry axis with respect to the solar equator, and aspect ratio. 
The quadrature position of the STEREO spacecraft with respect to those in the Sun-Earth line is advantageous to minimize uncertainties in the determination of the GCS parameters \citep[e.g.,][]{Cremades-etal2020}.

As anticipated in Section \ref{ss:obs}, the 3D latitude and longitude determined for the source regions and CMEs (columns 4, 5, 8, and 9 from Table \ref{tab:events}) are used to calculate deflection in the latitudinal and longitudinal directions (columns 11 and 12), as well as the total 3D deflection (last column of Table \ref{tab:events}). The kinematic and magnetic analysis is applied only to those events exhibiting a total 3D deflection $\Delta \Psi \gtrsim 20^{\circ}$.

\subsubsection{Tracking}
Although Table \ref{tab:events} lists the total deflection for each event, we are mostly interested in analyzing the spatio-temporal evolution of these deflections. We achieve this by tracking in time the 3D location of the erupting prominences and associated CMEs. To characterize the evolution of the prominence material we use the tie-pointing/triangulation reconstruction technique \citep[see e.g.,][]{inhester2006, mierla2008, mierla2009} on EUV images from SDO/AIA, STEREO/EUVI and PROBA2/SWAP. The method uses a pair of images to trace the line-of-sight of a specific point selected in one image into the FOV of the second image. This line is called the epipolar line \citep[see][for details on the epipolar geometry]{inhester2006}. The tie-pointing method is convenient when the triangulated structure is compact and well defined, as is the case of prominences. In particular, we try to apply this method to parcels of prominence material in the EUV low corona, that can later be tracked to a feature in the CME's core as detected in coronagraph images. 

\begin{figure}
 \centering
 \includegraphics[width=\textwidth]{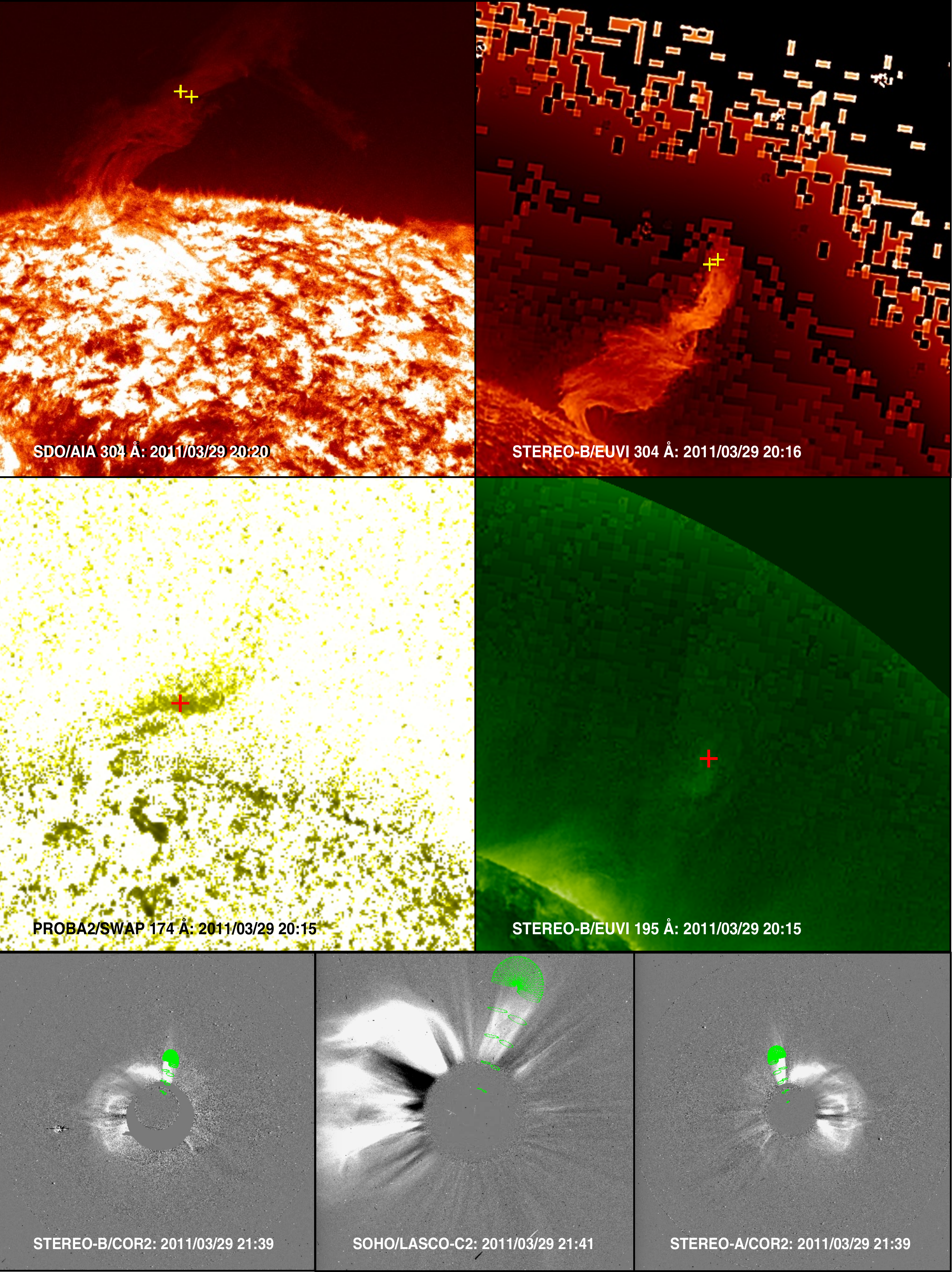}
 \caption{Top and middle panel: Triangulation of a parcel of the erupting prominence for the event of 29 March 2011. The top-left image corresponds to SDO/AIA 304\,\AA{} at 20:20 UT and the top-right to a wavelet-enhanced image of STEREO-B/EUVI 304\,\AA{} at 20:16 UT. In the middle panel the left image is a processed image of PROBA2/SWAP 174\,\AA{} and the right one a wavelet-enhanced STEREO-B/EUVI 195\,\AA{}, both at 20:15 UT. Yellow crosses indicate the parcel that is being triangulated to determine its 3D coordinates. Bottom panel: GCS model (green mesh) applied to the CME associated to the event on 29 March 2011. The left image corresponds to STEREO-B/COR2 at 21:39 UT, the central to SOHO/LASCO-C2 at 21:41 UT and the right one to STEREO-A/COR2 at 21:39 UT.}
 \label{fig:triang_gcs}
\end{figure}

In the top and middle panel of Figure~\ref{fig:triang_gcs} we show, for illustration purposes, two snapshots of the triangulation procedure for one of the events (29 March 2011) using SDO/AIA and STEREO-B/EUVI, both in 304\,\AA{} (top), PROBA2/SWAP 174\,\AA{} and STEREO-B/EUVI 195\,\AA{} (middle). The yellow crosses in each image indicate the parcel of filament that is triangulated to obtain its 3D coordinates.                                        
We triangulate a parcel of prominence until it either leaves the FOV of the EUV instruments or it becomes so faint that it cannot be further distinguished as a defined structure. For 4 events we triangulated the filament using PROBA2/SWAP images in 174\,\AA{} (for example the event showed in Figure~\ref{fig:triang_gcs}), taking advantage of its larger FOV, together with 195\,\AA{} images from STEREO/EUVI. We use EUVI 195\,\AA{} images instead of EUVI 171\,\AA{} because in general the cadence of 171\,\AA{} observations is very low (typically one image every 2 hours) compared to 195\,\AA{}, so the matching of these images with SWAP 174\,\AA{} is most of the times not possible. To perform measurements in a systematic way and because their maximum temperature responses are similar, we chose 195\,\AA{} to accomplish this task. Although the prominence may appear different in both wavelengths, the parcel of the prominence that is triangulated is usually located at the top of the structure and is easily recognizable as a bright feature against the dark background of the off-limb corona as the eruption progresses. For other studies using pairs of images in different wavelengths for the triangulation procedure please see \citet{seaton2011,mierla2013}. We also apply the triangulation technique to pairs of white-light images, whenever we can visually track the triangulated prominence parcel to the CME core seen by the coronagraphs.

To track the CME evolution, we implemented the GCS model at different time instants. The bottom panel of Figure~\ref{fig:triang_gcs} displays an example of the fitting for a time instant for 29 March 2011. We typically used image triplets from STEREO COR1 and COR2 in combination with SOHO/LASCO-C2, except for 2 cases in which we also used LASCO-C3 because the CME quickly leaves the LASCO-C2 FOV. The obtained GCS parameters of latitude, longitude, and height of the CME apex, added to those measured using the triangulation technique on the prominence are useful to analyze the spatio-temporal evolution of both structures.

\subsection{Magnetic energy density maps}
\label{ss:magnetic_field}

\begin{figure}
    \centering
    \includegraphics[width=0.8\textwidth]{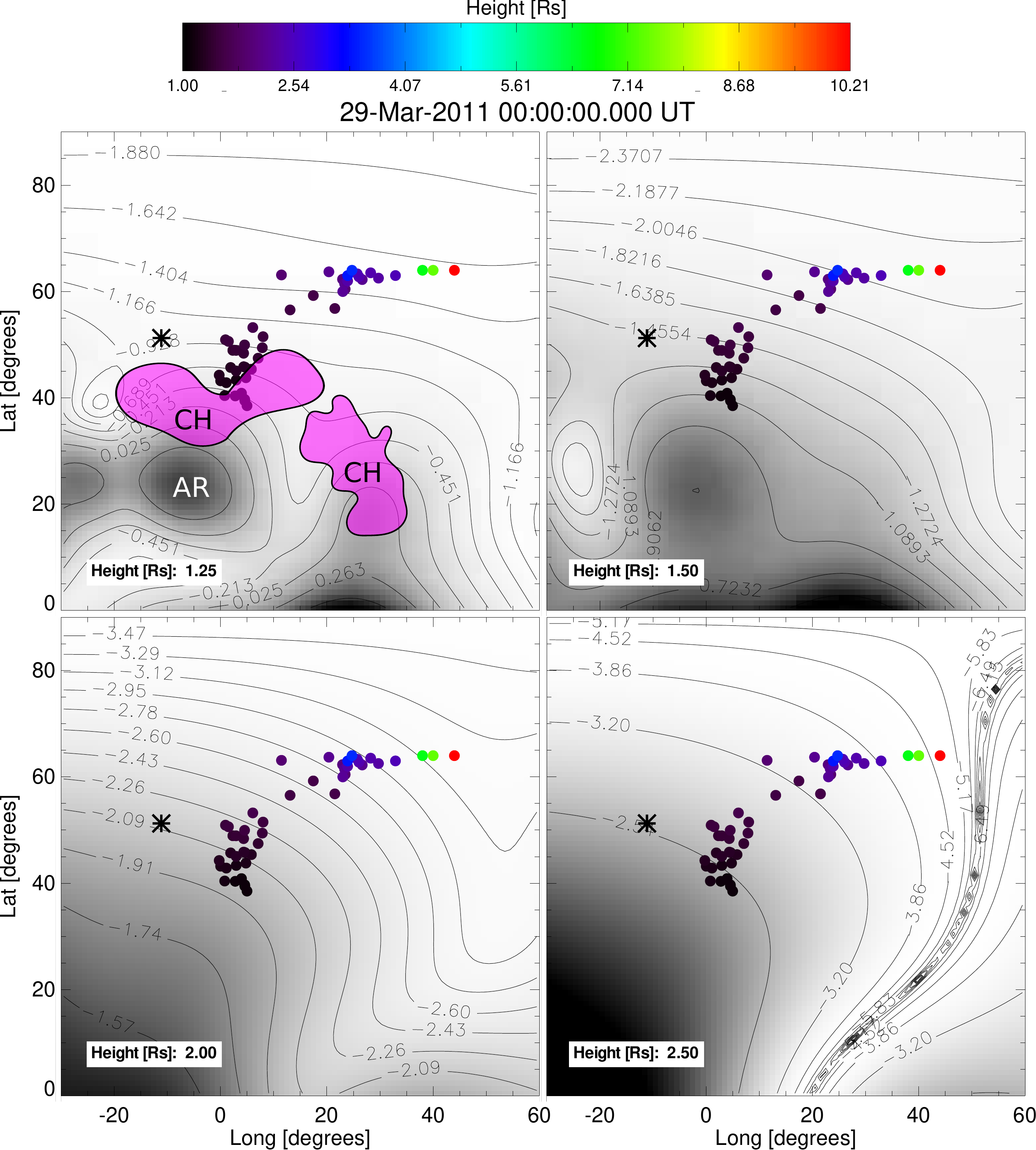}
    \caption{Magnetic energy density maps at different heights for the event occurred on 29 March 2011. The corresponding height is indicated in the bottom left corner of each panel. The gray scale represents the intensity of the magnetic energy, being the darker regions where the intensity is higher. The contours (solid black lines) indicate also the magnetic energy in logarithmic scale. In the map corresponding at $2.5\,R_{\odot}$ the HCS is delimited by a thick black curve. The magenta filled contours denote the coronal holes (CH) obtained from EUV images and the active regions are pointed with AR. The filled circles show the coordinates obtained from the measurements using tie-pointing for the prominence and GCS model for the CME. The color of each circle indicates its height according to the rainbow scale in the top.}
    \label{fig:mag_vs_h}
\end{figure}

To analyze the relationship between prominence/CME deflection and the magnetic environment, i.e., how the surrounding coronal conditions affect the trajectory of both structures, we compute maps of energy density associated to the magnetic field ($B$). The magnetic energy density ($\propto B^2$) distribution at different heights is determined from the potential field  source surface (PFSS) model by \citet{schrijver2003}. This model is valid in a height range between $1\,R_{\odot}$ and $2.5\,R_{\odot}$. Figure~\ref{fig:mag_vs_h} shows example of magnetic energy density maps for 29 March 2011 at different heights. The iso-contours (in logarithmic scale) overplotted on top of the (gradient-filled) gray background indicate levels of constant $B^2$ (as indicated by the iso-contours, darker regions correspond to higher magnetic energy values). The black asterisk represents the central position of the source region and the circles indicate the triangulated prominence points and the ones obtained from the GCS model of the CME. The color of the circles indicates the height. At lower heights (top panels) we can see localized structures as active regions (AR), to the south of the measured points, and two coronal holes (CH), also to the south. As the height increases (bottom panels), the global structure of the magnetic field becomes evident including the HCS. Note from the contour levels, that the intensity of the magnetic field decays at least two orders of magnitude within the considered height range. PFSS 3D extrapolations are also used to examine the global magnetic field and search for the presence of magnetic structures such as coronal holes, helmet streamers and/or pseudostreamers in the vicinity of each source, erupting prominence, and CME.  

The magnetic energy density maps enables the estimation of the local magnetic gradient, to determine the possible influence of the magnetic field in the trajectory of the prominences and CMEs.

\subsection{Trajectory in the $\Theta-\Phi$ plane and gradient of magnetic energy density} 
\label{ss:magnetic_deflection}

\begin{figure}
    \centering
    \includegraphics[angle=90,width=0.6\textwidth]{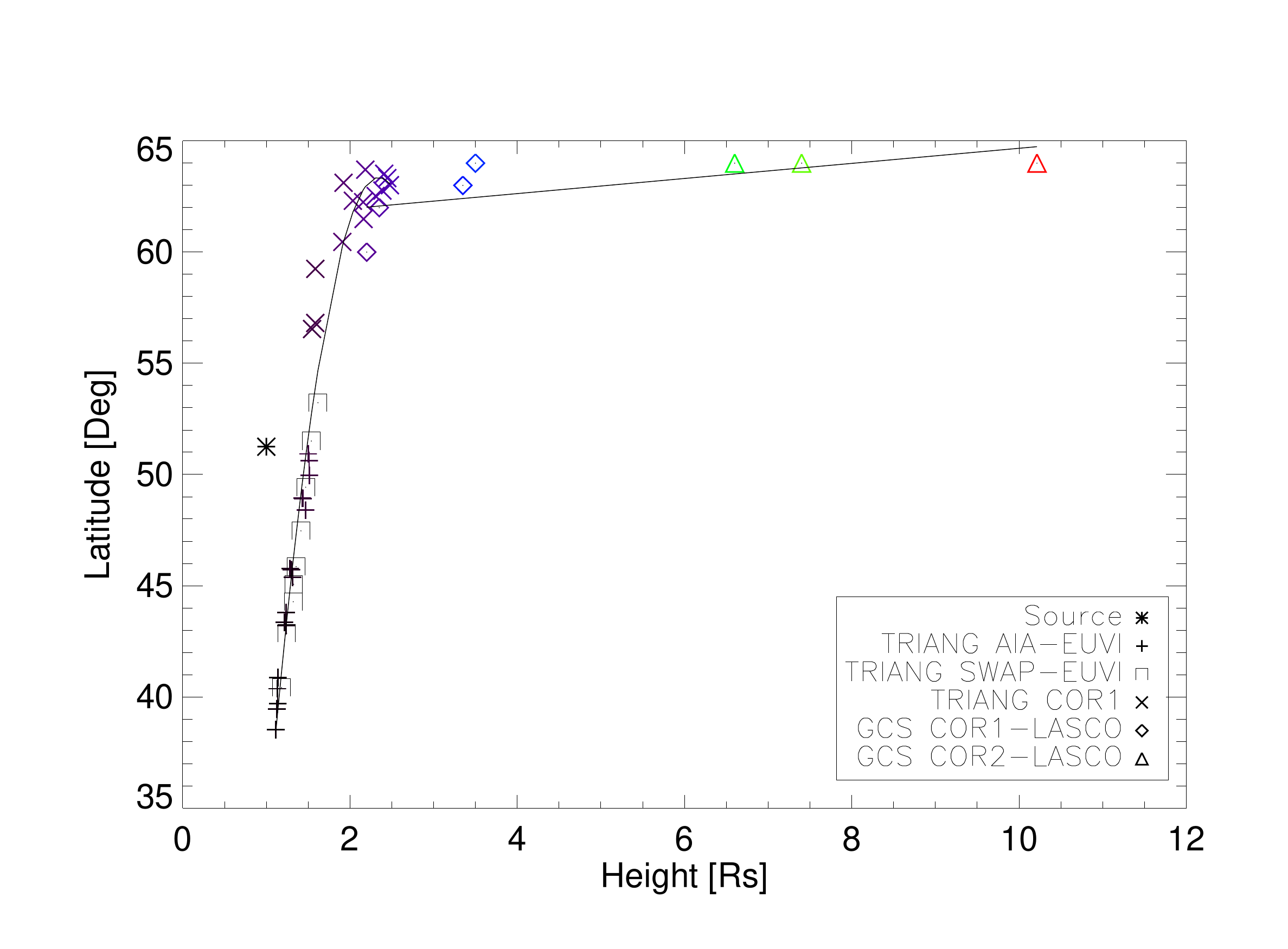}
    \includegraphics[angle=90,width=0.6\textwidth]{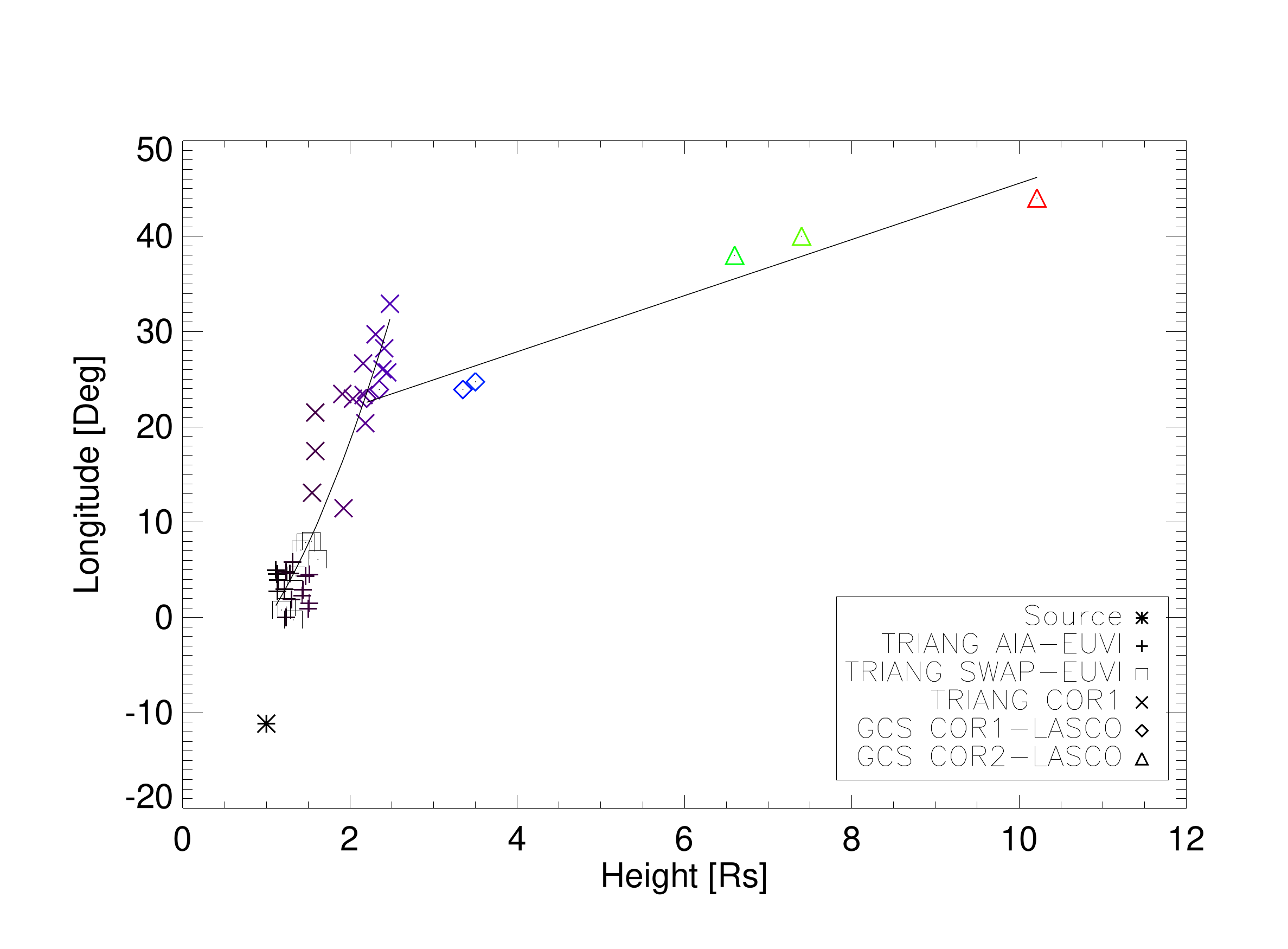}
    \includegraphics[angle=90,width=0.6\textwidth]{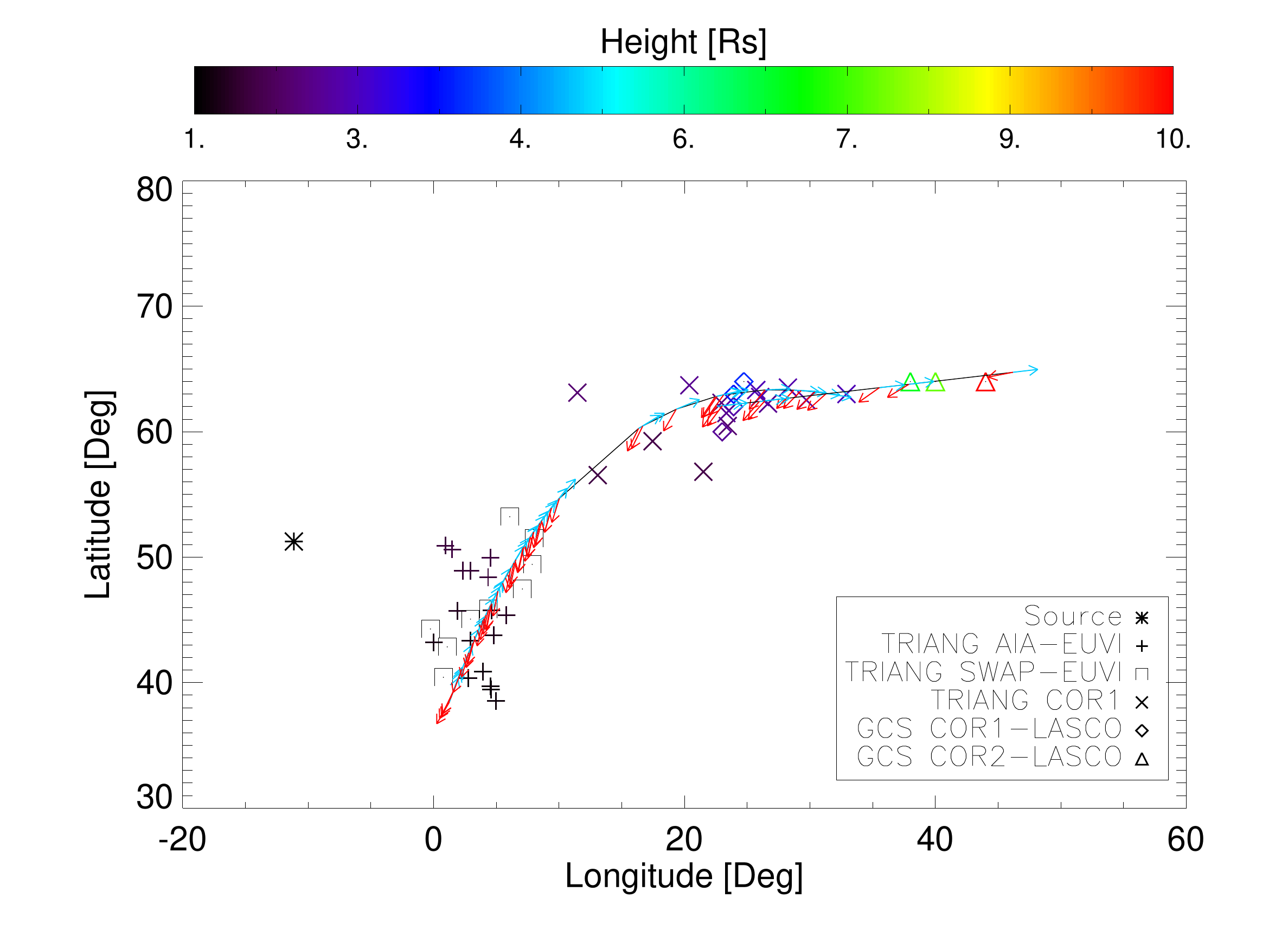}
    \caption{Top and middle panels: Latitude and longitude, respectively, as a function of height for the event on 29 March 2011. The various symbols indicate the measurements of the coordinates using different methods and imagers. The solid black lines correspond to a quadratic fit applied to the prominence data and a linear fit applied to the CME data series. Bottom panel: Trajectories (black solid lines) projected on the $\Theta-\Phi$ plane resulting from the fitted curves. Cyan arrows represent the direction of the tangent vector and red ones show the direction of the magnetic energy density gradient. The color scale of the measured points indicates their height.}
    \label{fig:tangent}
\end{figure}

Here, we examine the effect of the magnetic field in deflecting the investigated structures, i.e., both erupting prominence and CME, by analyzing their 3D trajectory in the context of the magnetic configuration, which is provided by magnetic energy density maps. From the variability of latitude and longitude with time and height, it is possible to plot the trajectory projected in the plane latitude vs. longitude ($\Theta-\Phi$ plane). As a first step, we plot latitude and longitude as a function of height, as in the example displayed in the top panels of Figure~\ref{fig:tangent}. The different symbols are measurements resulting from the various instruments, while their color coding represents height. Data series ``TRIANG AIA--EUVI'' and ``TRIANG SWAP--EUVI'' denote triangulations of prominence parcels performed in the low corona. Additionally, note that the data series ``TRIANG COR1'' corresponds to parcels of the prominence identified in the CME core and tracked in the COR1 A and B coronagraphs; whereas ``GCS'' data series refer to the CME apex. Solid lines are quadratic and linear fits to the latitude and longitude coordinates obtained for the filament and the CME. We use linear or quadratic functions according to the behavior of the prominence and CME for each event. We don't include the source in the prominence fit because this measurement corresponds to a different part of the triangulated filament. 

The bottom panel of Figure~\ref{fig:tangent} displays the resulting trajectory of both data series projected onto the $\Theta-\Phi$ plane. Vectors tangent to the curve, described by $\frac{d \Theta}{d \Phi}= \frac{d \Theta/dh}{d\Phi/dh}$, are plotted as cyan arrows for several points over the fitted trajectory. At the location of these points we also calculate the direction of the gradient of magnetic energy density computed from the magnetic density maps, see Fig.~\ref{fig:mag_vs_h}. The direction of the magnetic gradient is displayed with red arrows. It can be assumed that the magnetic field becomes predominantly radial for heights above $2.5\,R_{\odot}$, in which case the magnetic energy density would basically change only in the radial direction, and not in the $\Theta-\Phi$ plane. Therefore, for heights above $2.5\,R_{\odot}$ gradients are assumed to keep the same direction. The length of the cyan and red arrows are scaled to have comparable sizes for visualization purposes, hence they do not represent the actual magnitude of the tangent and the magnetic gradient. To quantify whether the trajectory is aligned with the direction of the magnetic gradient, we determine the angle between these two vectors. These results are shown in Section \ref{ss:magnetic_influence}. 

\section{Results}
\label{s:results}

With the aim of performing a systematic study of CMEs having large deflections, we focus the analysis on the main sources of deflection previously studied by another authors \citep[e.g.,][]{gui2011,liewer2015,kay2015}: the influence of the magnetic force and the kinematic features of both structures, prominence and associated CME.

\subsection{The role of the magnetic environment on deflection}
\label{ss:magnetic_influence}

The measured coordinates of source region, prominence parcels, and CME apex plotted as symbols against synoptic maps of magnetic energy density (built as explained in Section \ref{ss:magnetic_field}), allows to comprehensively visualize the location of the various structures. Given their significance, in Figures~\ref{fig:densmaps1} and \ref{fig:densmaps2} we show all resulting plots for the 13 analyzed events considering the synoptic maps at $2.5\,R_{\odot}$. The gray background shows the intensity of the magnetic energy, with darker regions having the highest  magnetic energy and brighter regions associated with lower magnetic energy. The HCS is indicated with a thick solid black line and the other solid black lines represent contour levels of the magnetic energy. The  reconstructed points are displayed as colored circles, with black representing the lowest height ($1\,R_{\odot}$) and red the greatest ($15\,R_{\odot}$) of all events. The source is indicated with a black asterisk. It can be appreciated how trajectories evolve in some cases by moving away from regions of high magnetic energy density and in other cases heading towards regions of low magnetic energy density. A quantitative way of evaluating such a behavior can be achieved by determining the angle between the tangent direction to the trajectory and the gradient of magnetic energy density, as described in Section \ref{ss:magnetic_deflection}. Henceforth we will call this angle $\delta$. Ideally, ejecta moving directly towards the HCS or a local minimum energy region and away from high magnetic energy regions, i.e. exactly against the gradient of magnetic energy density, would present an angle $\delta \sim180^\circ$.
 \begin{figure}
     \centering
     \includegraphics[width=0.07\textwidth, angle=90, trim=15cm 0.5cm 0.5cm 0.5cm, clip=true]{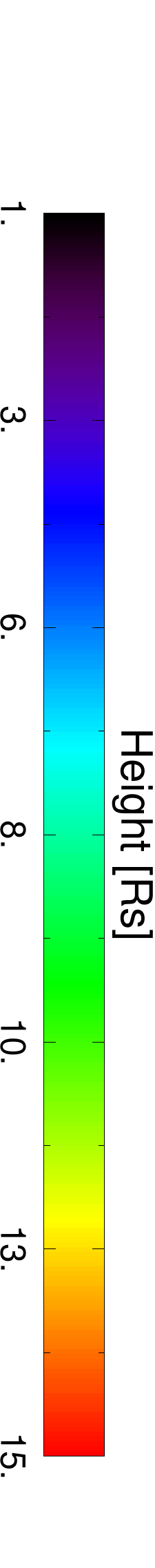}
     \includegraphics[width=0.28\textwidth,angle=90, trim=3.5cm 3.cm 4.0cm 2.5cm, clip=true]{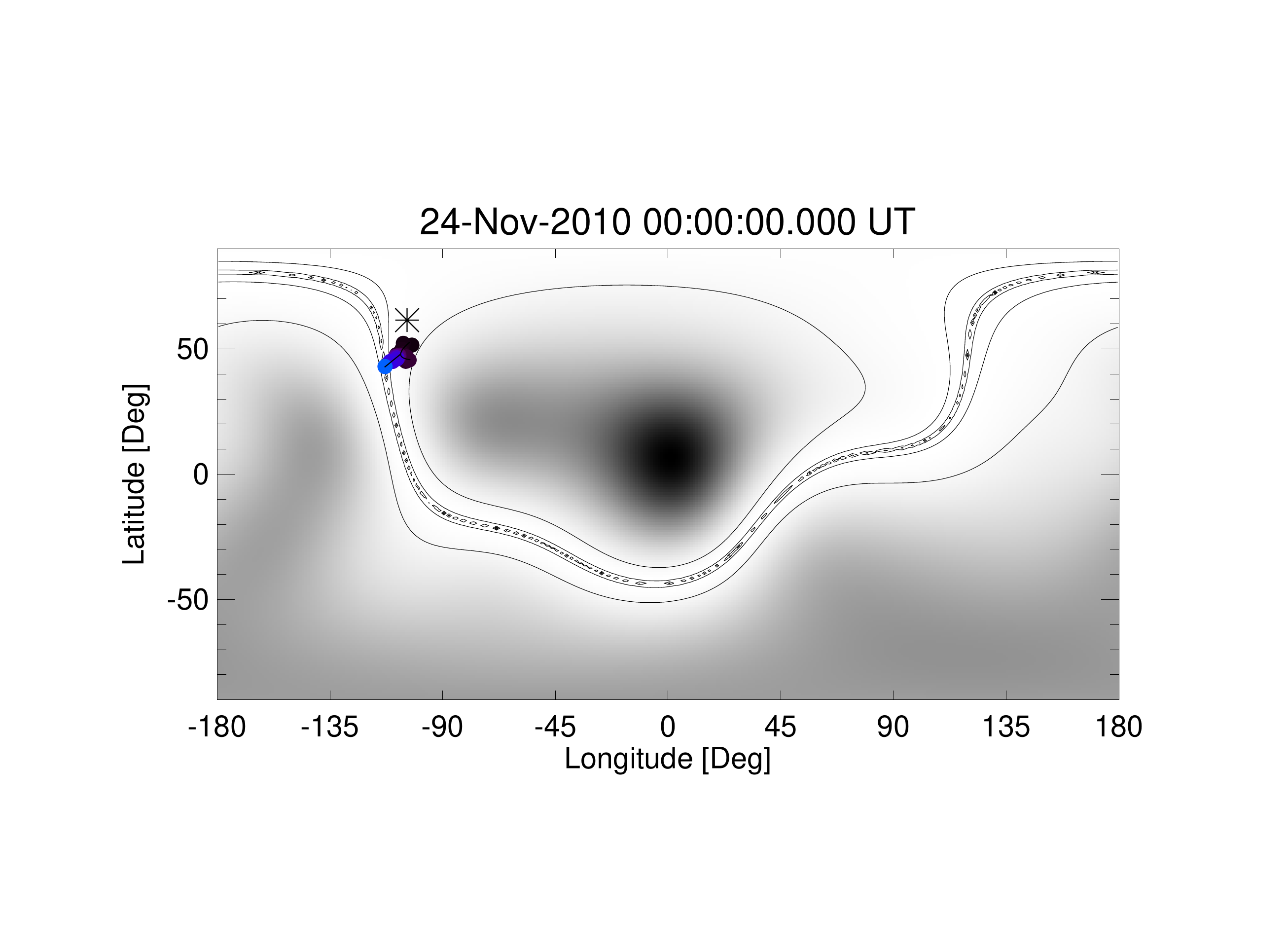}
     \includegraphics[width=0.28\textwidth,angle=90, trim=3.5cm 3.cm 4.0cm 2.5cm, clip=true]{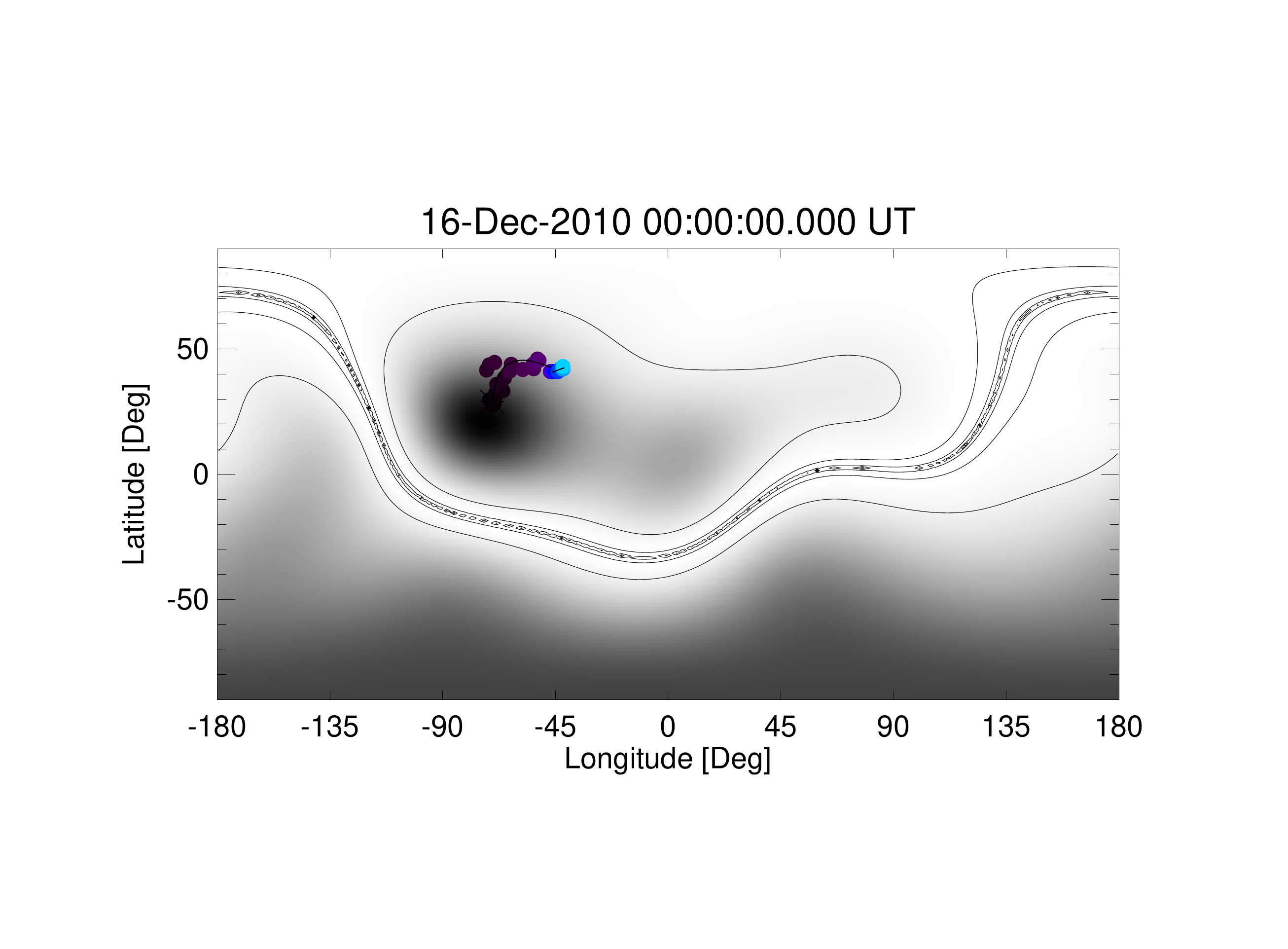}
     \includegraphics[width=0.28\textwidth,angle=90, trim=3.5cm 3.cm 4.0cm 2.5cm,  clip=true]{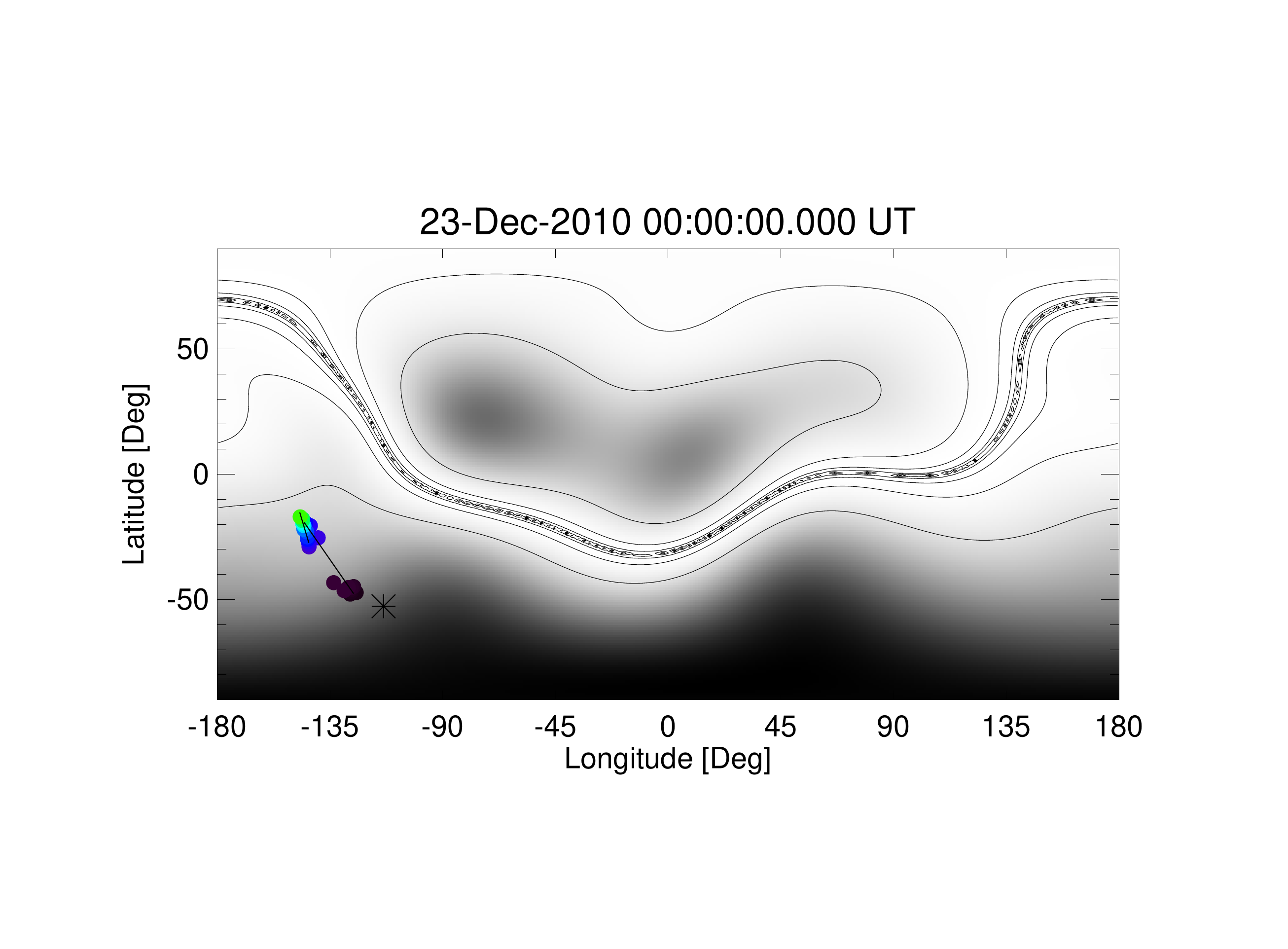}
     \includegraphics[width=0.28\textwidth,angle=90, trim=3.5cm 3.cm 4.0cm 2.5cm, clip=true]{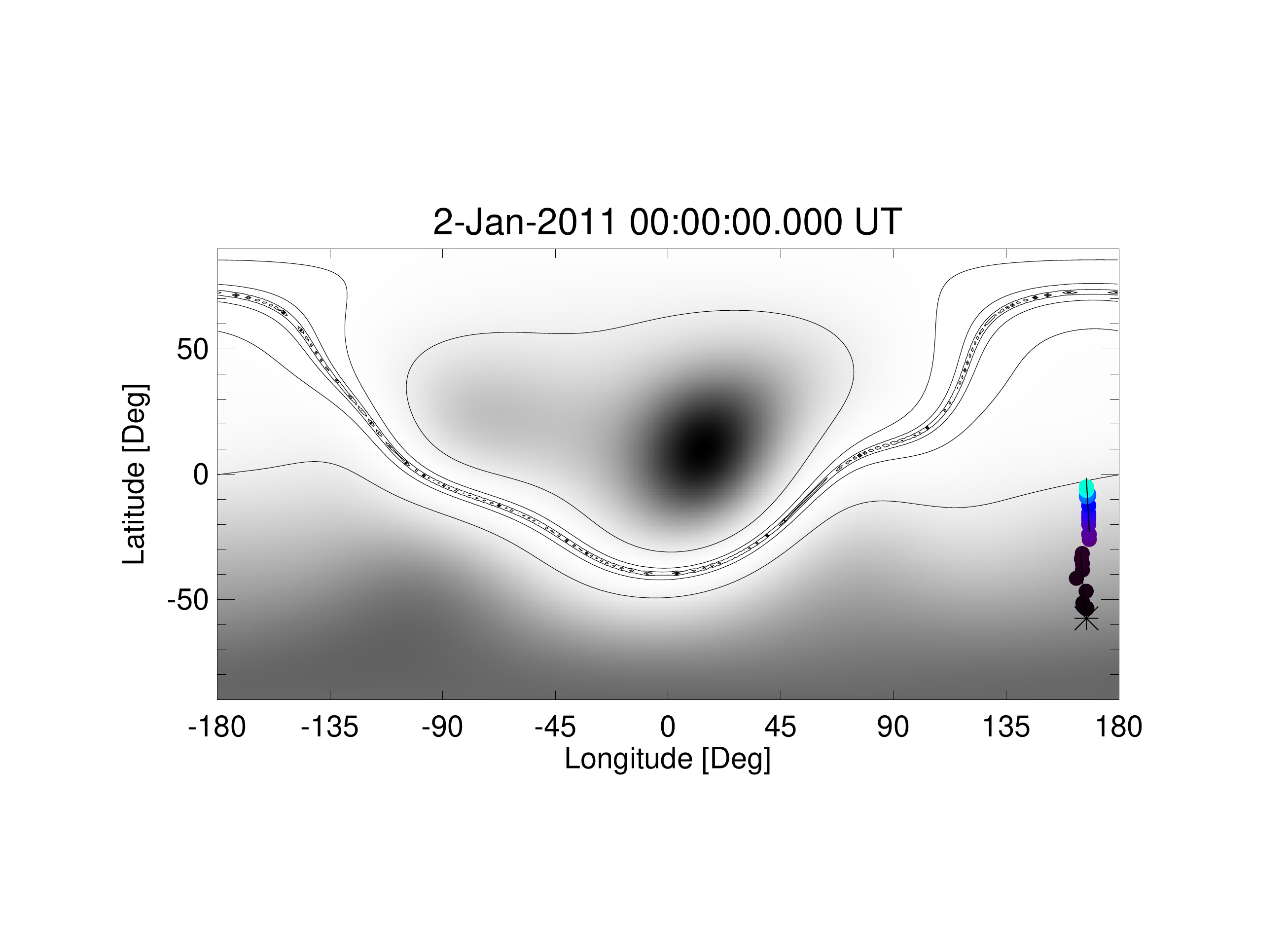}
     \includegraphics[width=0.28\textwidth,angle=90, trim=3.5cm 3.cm 4.0cm 2.5cm, clip=true]{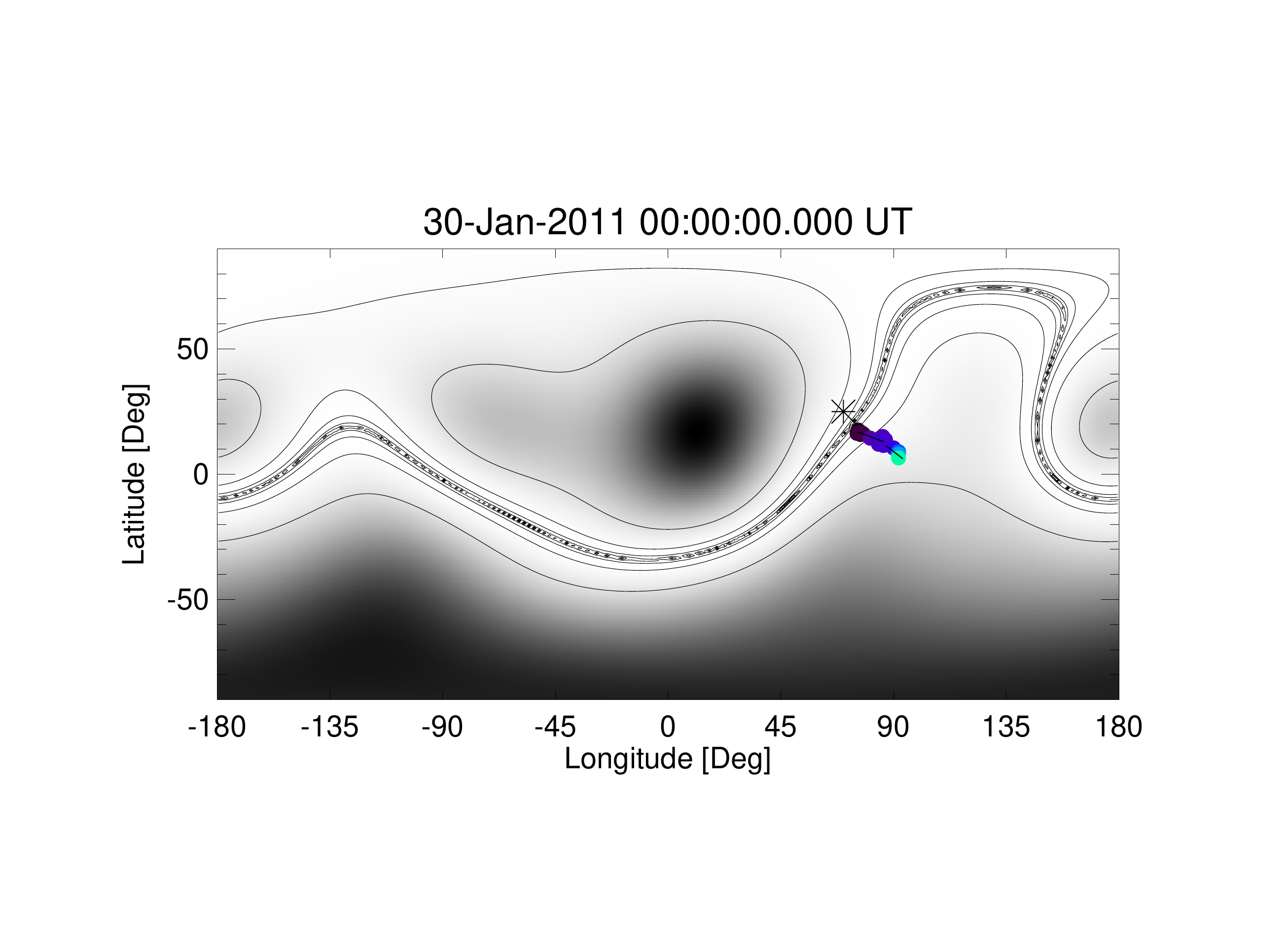}
     \includegraphics[width=0.28\textwidth,angle=90, trim=3.5cm 3.cm 4.0cm 2.5cm, clip=true]{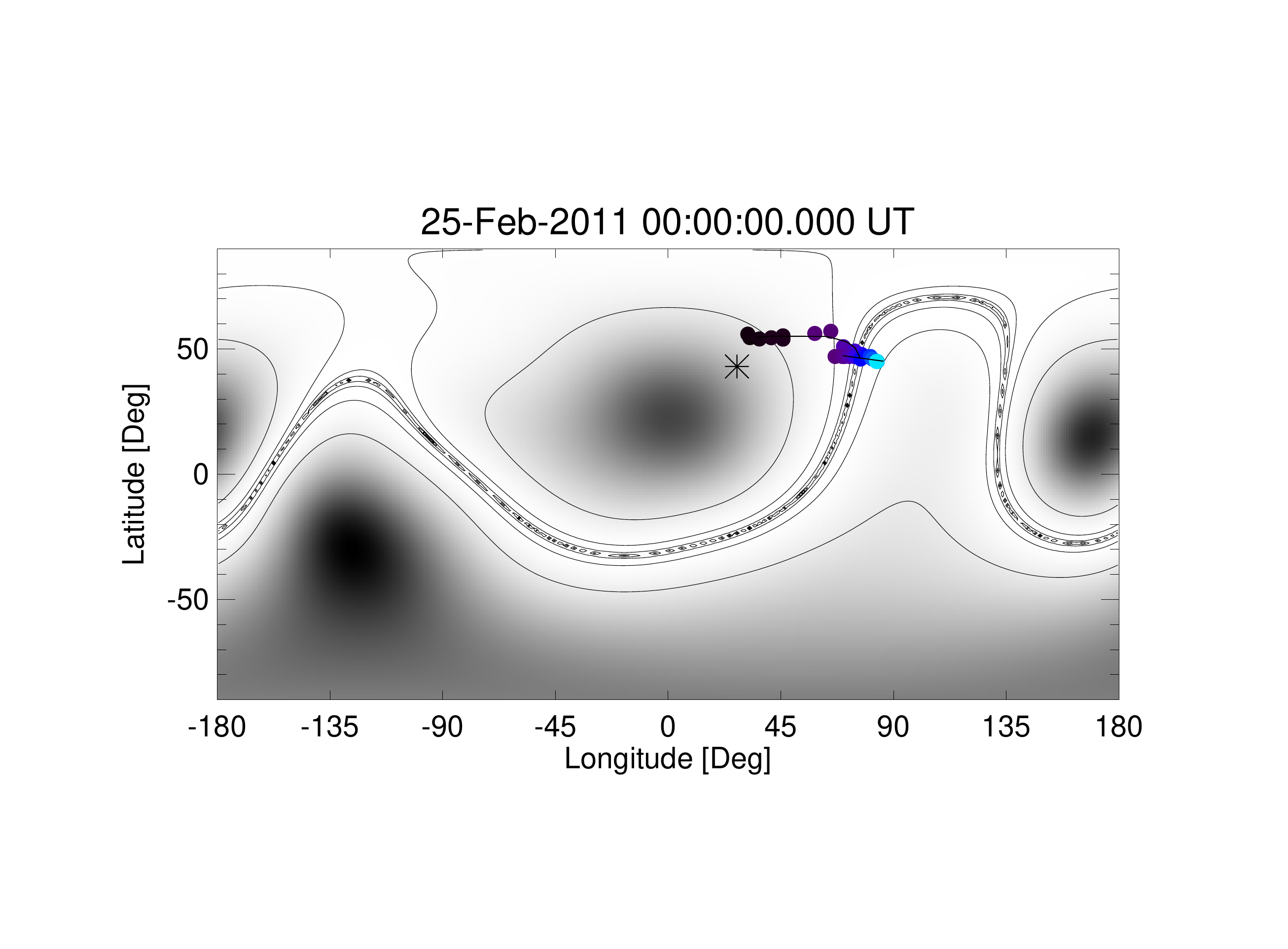}
     \includegraphics[width=0.28\textwidth,angle=90, trim=3.5cm 3.cm 4.0cm 2.5cm, clip=true]{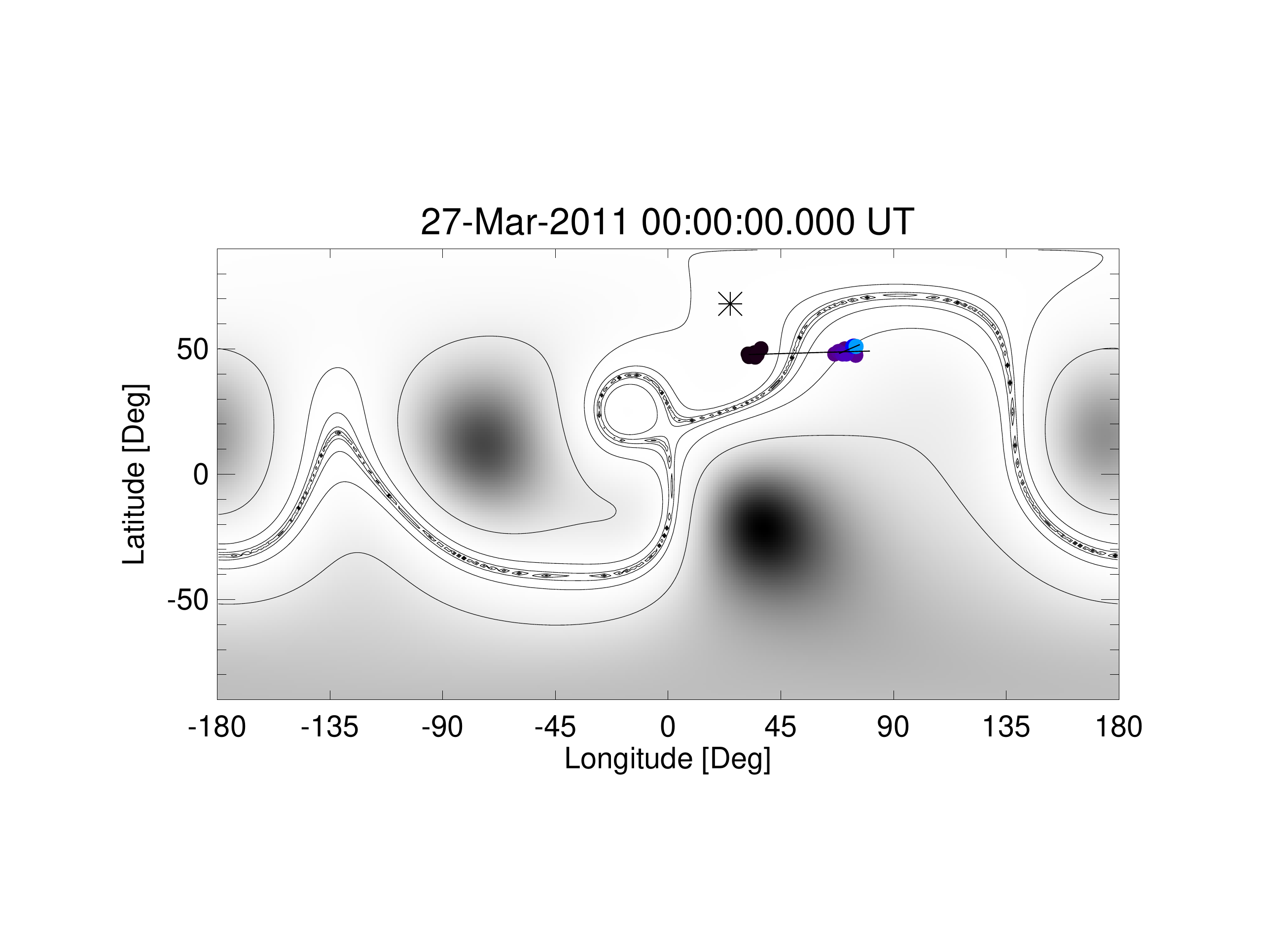}
     \includegraphics[width=0.28\textwidth,angle=90, trim=3.5cm 3.cm 4.0cm 2.5cm, clip=true]{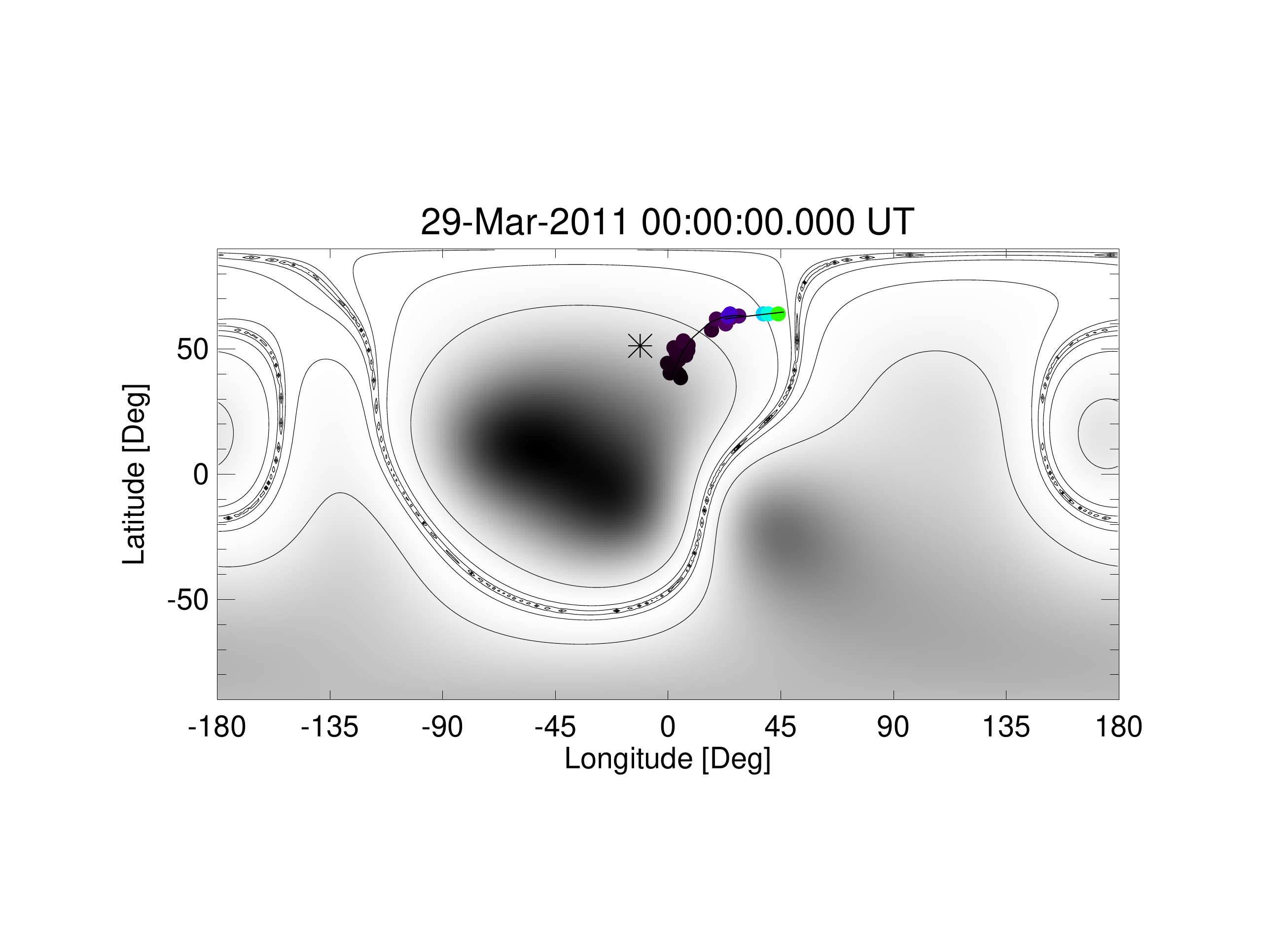}
     \includegraphics[width=0.28\textwidth,angle=90, trim=3.5cm 3.cm 4.0cm 2.5cm, clip=true]{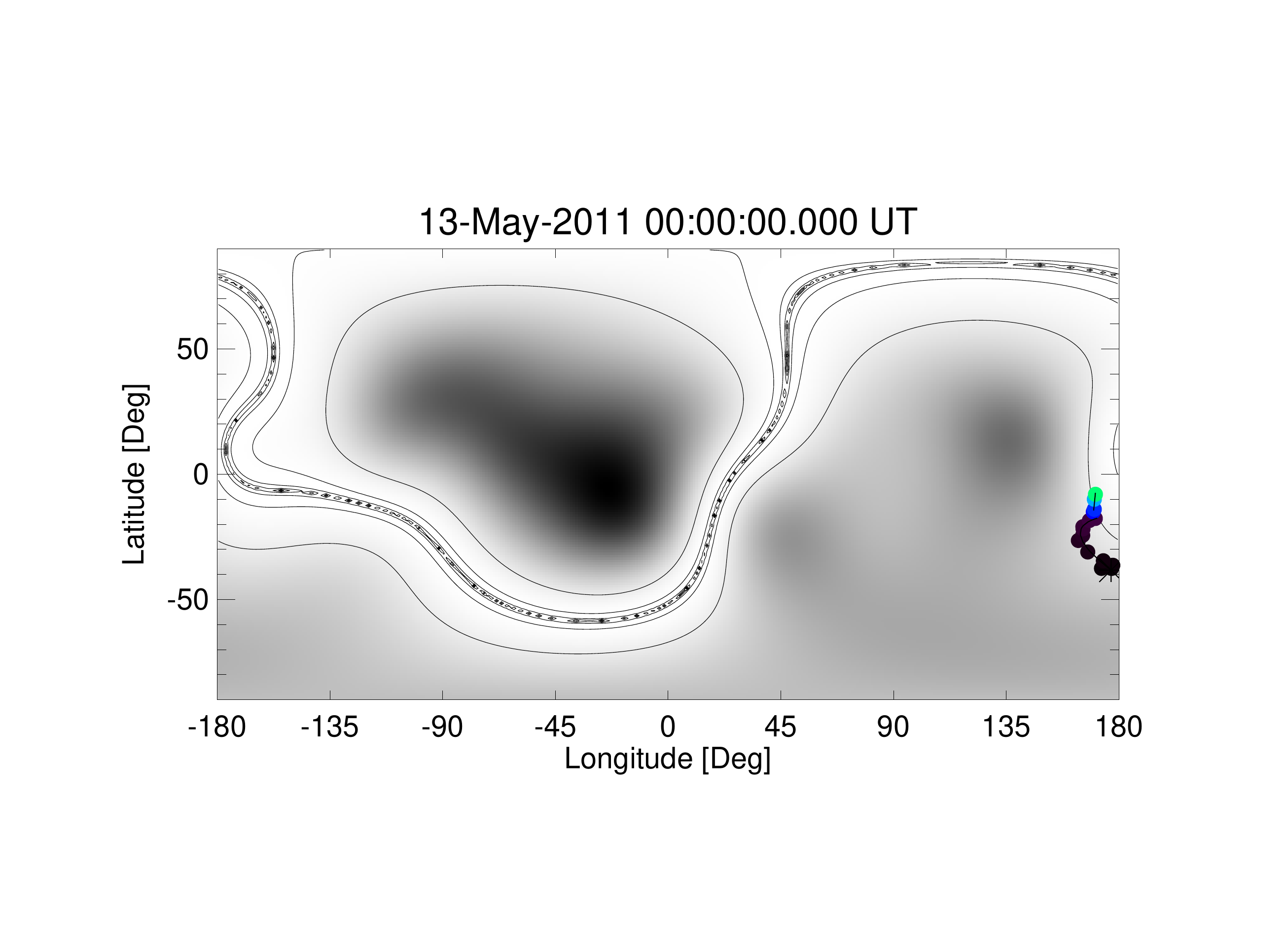}
     \includegraphics[width=0.28\textwidth,angle=90, trim=3.5cm 3.cm 4.0cm 2.5cm, clip=true]{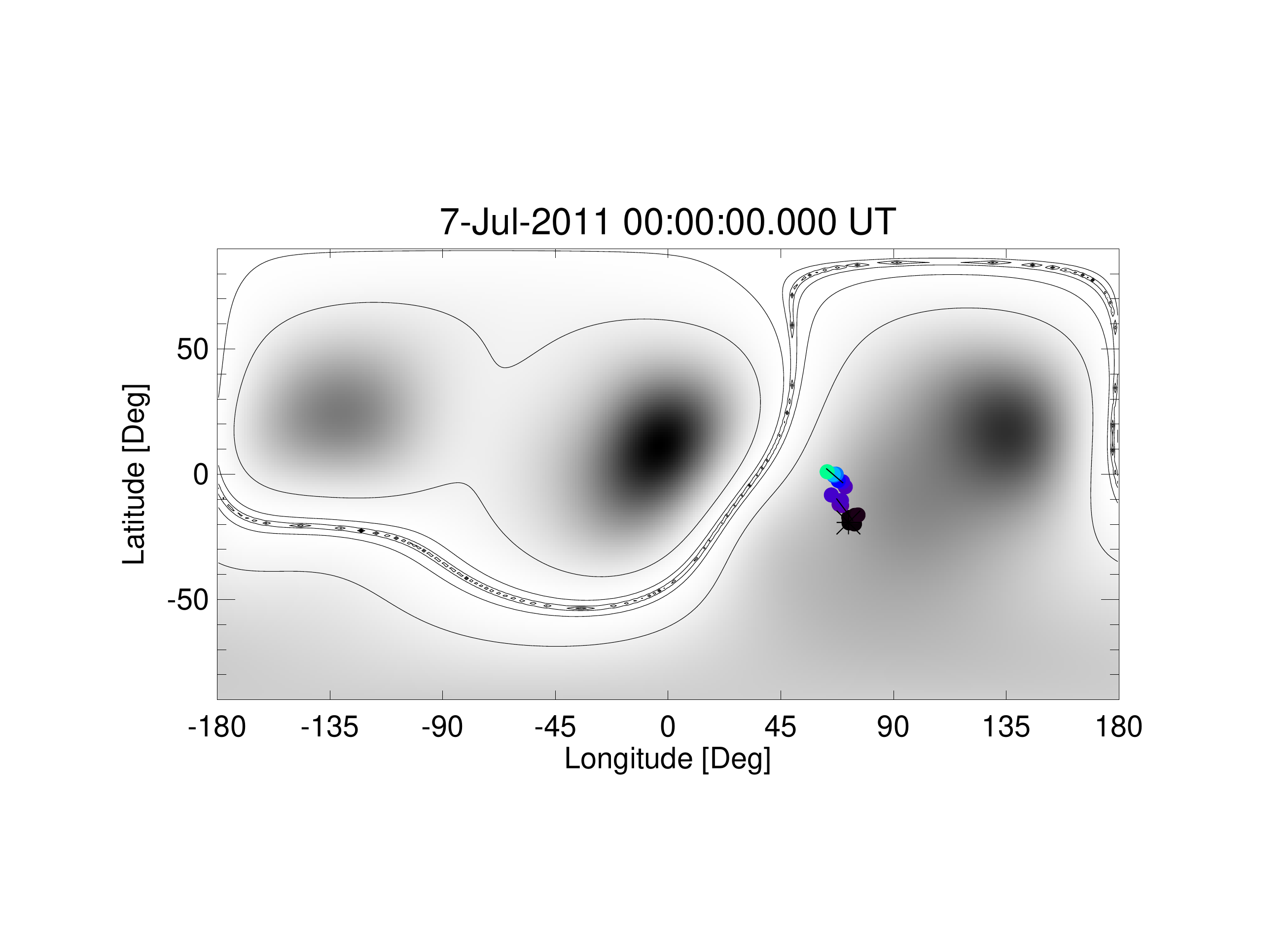}
     \caption{Synoptic maps of magnetic energy density (grey-scale shaded background) at 2.5\,$R_\odot$ for the dates of the 13 events in Table \ref{tab:events}. Solid lines are contours of low magnetic energy density. The thick solid black line indicates the HCS. The colored dots represent the coordinates of tracked prominence parcels and CME apex, with the color coding representing their height. The dots are connected by lines that represent the fitted trajectory. The source region is indicated with a black asterisk.}
     \label{fig:densmaps1}
 \end{figure}
 \begin{figure}
     \centering
     \includegraphics[width=0.28\textwidth,angle=90, trim=3.5cm 3.cm 4.0cm 2.5cm, clip=true]{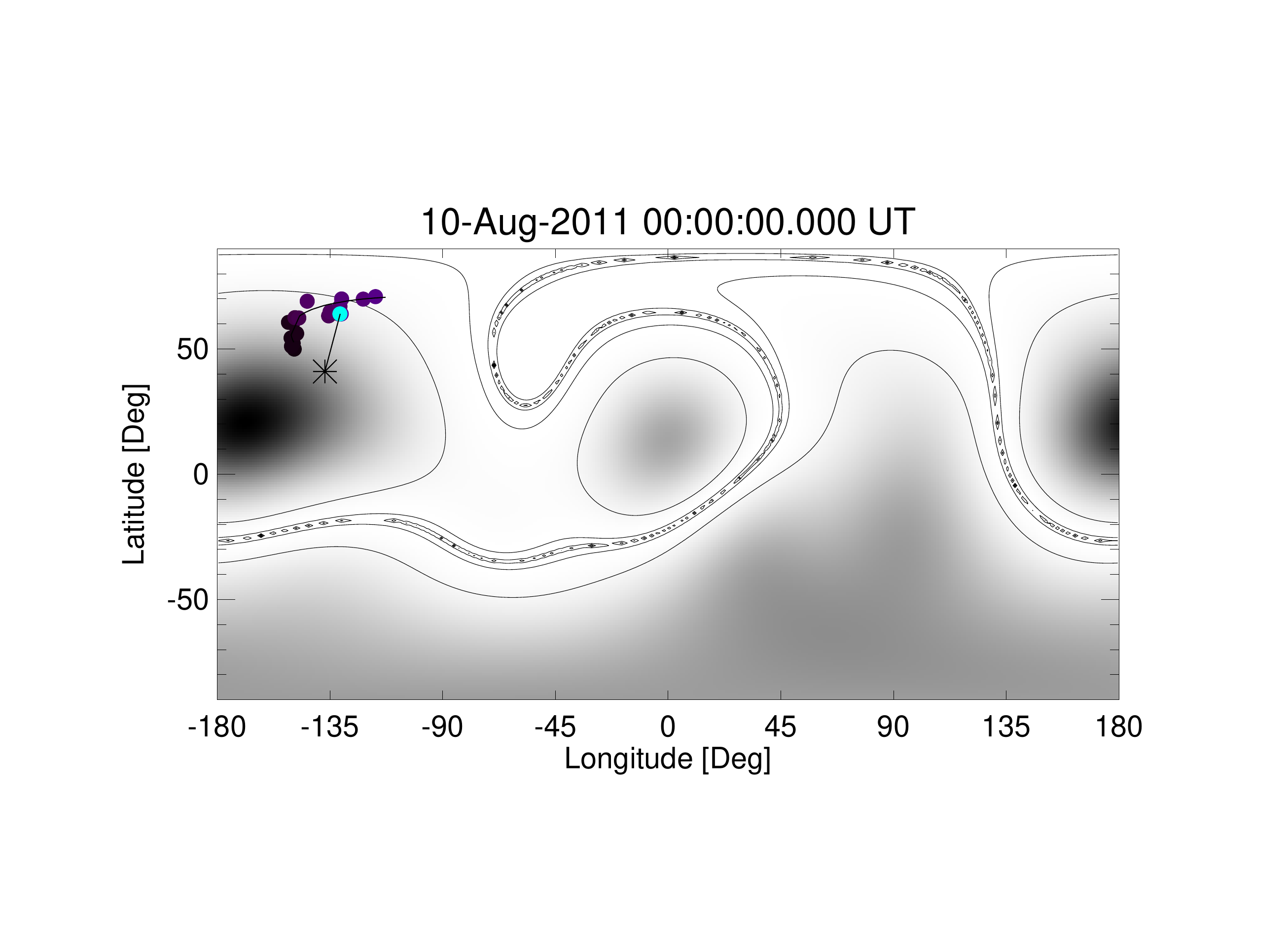}
     \includegraphics[width=0.28\textwidth,angle=90, trim=3.5cm 3.cm 4.0cm 2.5cm, clip=true]{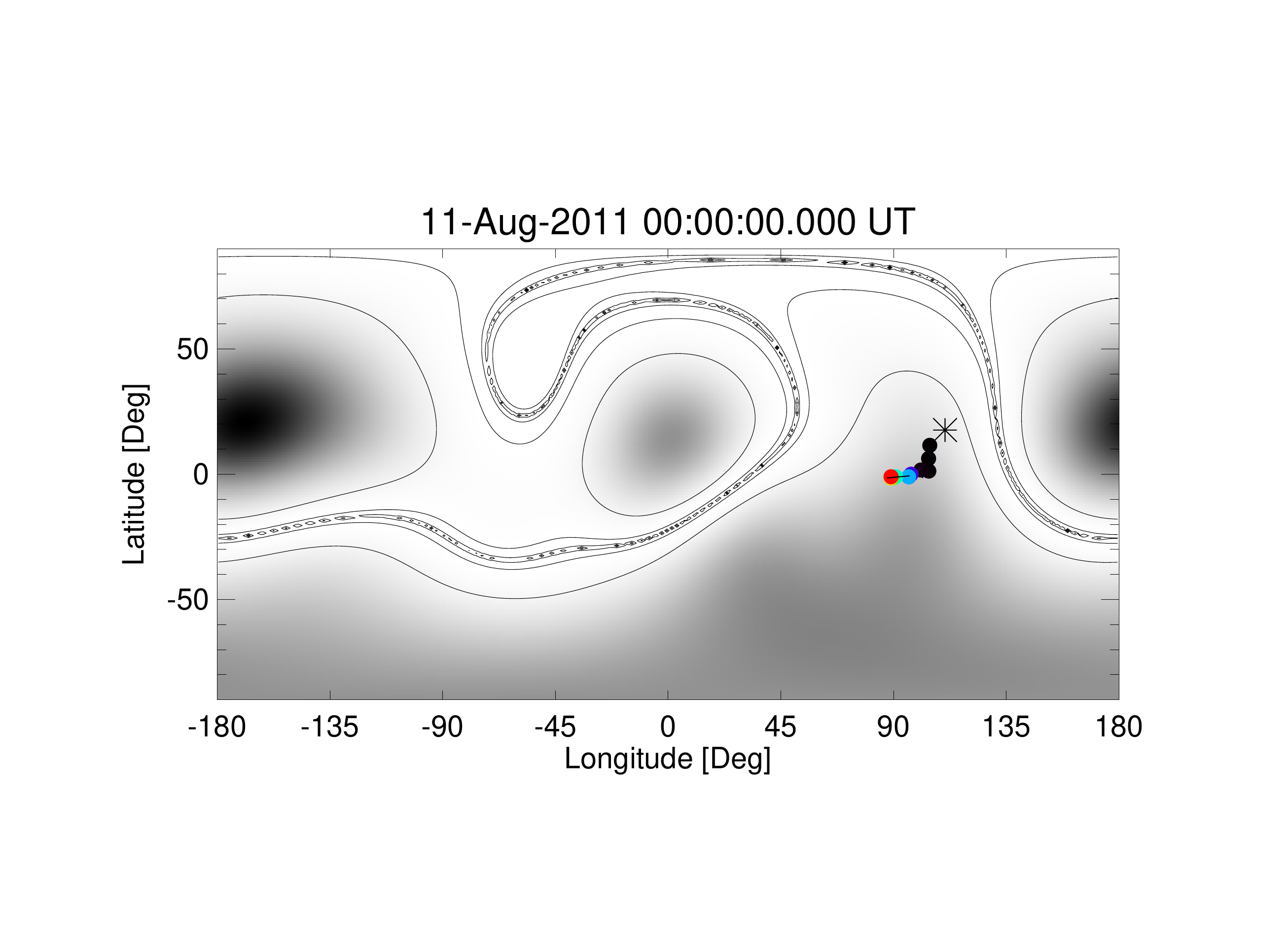}
     \includegraphics[width=0.28\textwidth,angle=90, trim=3.5cm 3.cm 4.0cm 2.5cm, clip=true]{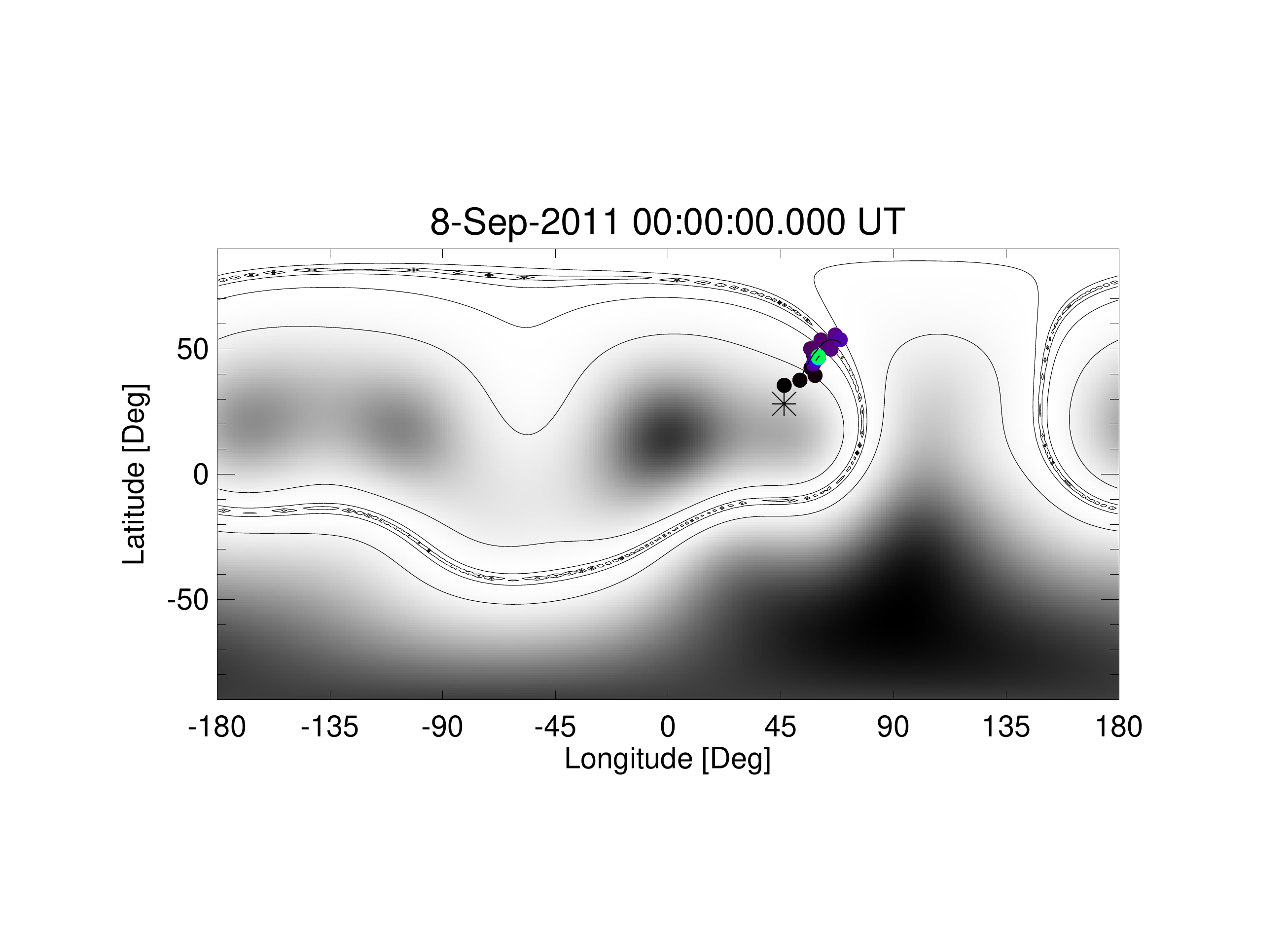}
     \caption{(cont.) Idem Figure \ref{fig:densmaps1}.}
     \label{fig:densmaps2}
 \end{figure}

\begin{figure}
    \centering
    \includegraphics[width=0.45\textwidth, angle=90]{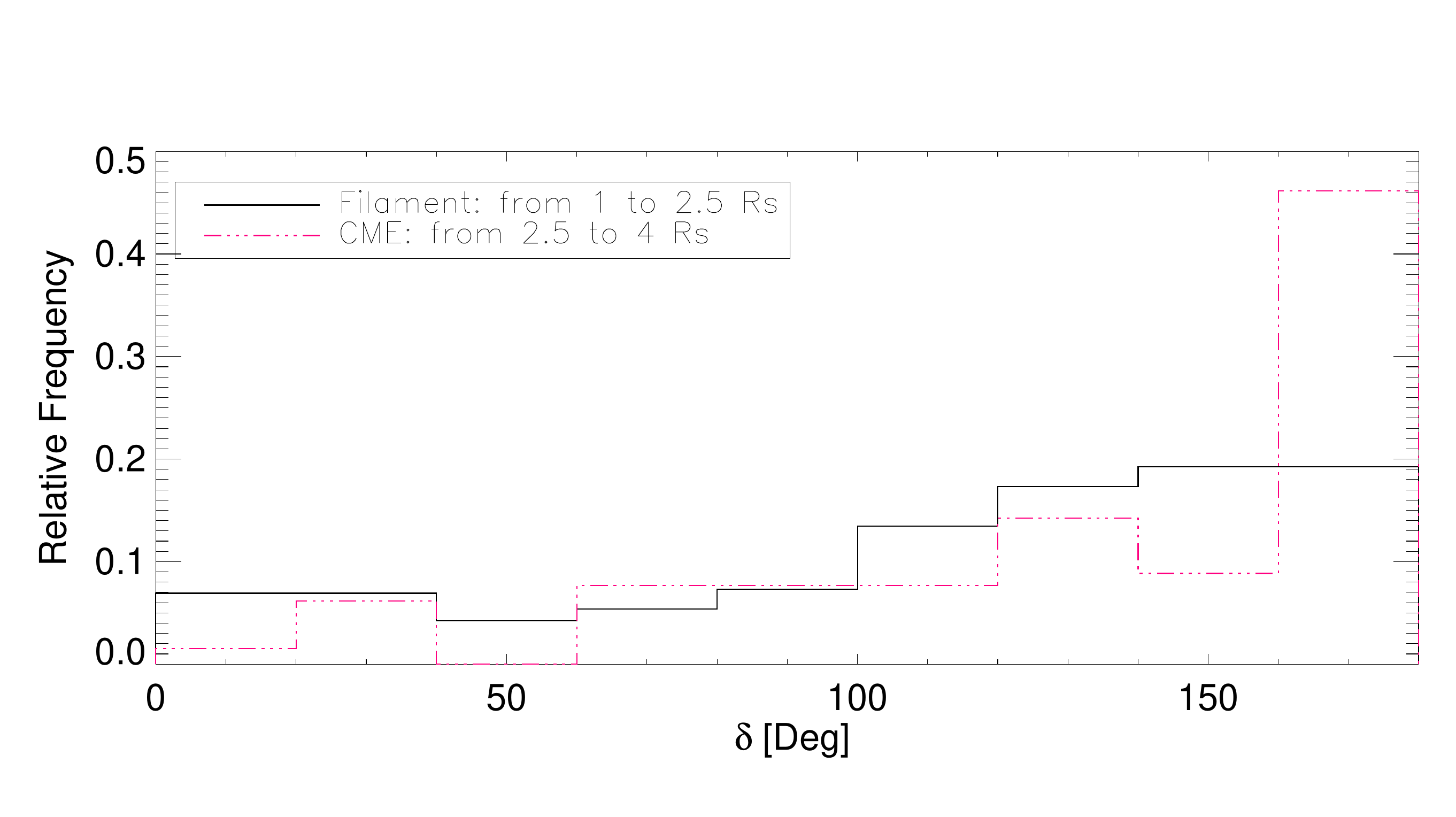}
    \caption{Distribution of the $\delta$ (angle between the trajectory tangent and the gradient of the magnetic energy) corresponding to the two analyzed structures. The distribution of the angle for filament parcels is shown with a black solid line considering the fitted curves from 1 to $2.5\,R_{\odot}$. For the CME this angle is considered between 2.5 to $4\,R_{\odot}$ and is shown with a magenta dot-dashed line.}
    \label{fig:delta_angle}
\end{figure}

In Figure \ref{fig:delta_angle} we show the distribution of $\delta$ as determined from each measured point of all events. The results obtained from filament parcels from 1 to 2.5$\,R_{\odot}$ are shown with a black solid line, while the magenta dashed line represents the angle distribution for CME measurements from 2.5 to 4$\,R_{\odot}$. For lower heights ($< 2.5\,R_{\odot}$) $\delta$ shows a flattened distribution, with 55\% of the values distributed between 120$^\circ$ and 180$^\circ$, while 25\% present values between 60$^\circ$ and 120$^\circ$ and the remaining 20\% show smaller angles.  The first group is related to erupting filaments located near ARs and CHs, the second and third group are related to regions of open magnetic field lines and quiet sun. For greater heights, between 2.5 to $4\,R_{\odot}$, the $\delta$ distribution is less disperse. Approximately 69\% of the values are between 120$^\circ$ and 180$^\circ$ (of which 46\% are concentrated between 160$^\circ$ and 180$^\circ$), 24\% present values between 60$^\circ$ and 120$^\circ$ and the remaining 7\% shows lower angles. This distribution indicates that the direction of the trajectory is mainly opposite to the direction of maximum magnetic energy growth, in agreement with previous reports \citep[e.g.,][]{gui2011}. The CMEs with $\delta$ between 120$^\circ$ and 180$^\circ$ leave the low corona near the HCS or a region of low magnetic energy, moving away from CHs. Some few CMEs that have $\delta$ angles $<120^\circ$ move beyond the HCS. All these cases are described in detailed in Section \ref{ss:anomalous_cases}. 

From Figure~\ref{fig:delta_angle}, it can also be noted that the filament distribution is more flatter than the CME distribution. In general, it can be said that the values of $\delta$ for altitudes $<2.5\,R_{\odot}$ fluctuate more with height than the values for $>2.5\,R_{\odot}$. This suggests that the alignment of the direction of deflection with the direction in which the magnetic energy decreases takes place more often at higher altitudes ($>2.5\,R_{\odot}$). 
To gain further insight into the properties of CME deflections, we performed a kinematic study of prominences and CMEs described in the following section. 

\subsection{Kinematic analysis}
\label{ss:kinematic}

\begin{figure}
    \centering
    \includegraphics[width=0.33\textwidth,angle=90, trim=1.0cm 1.5cm 2.0cm 1.0cm, clip=true]{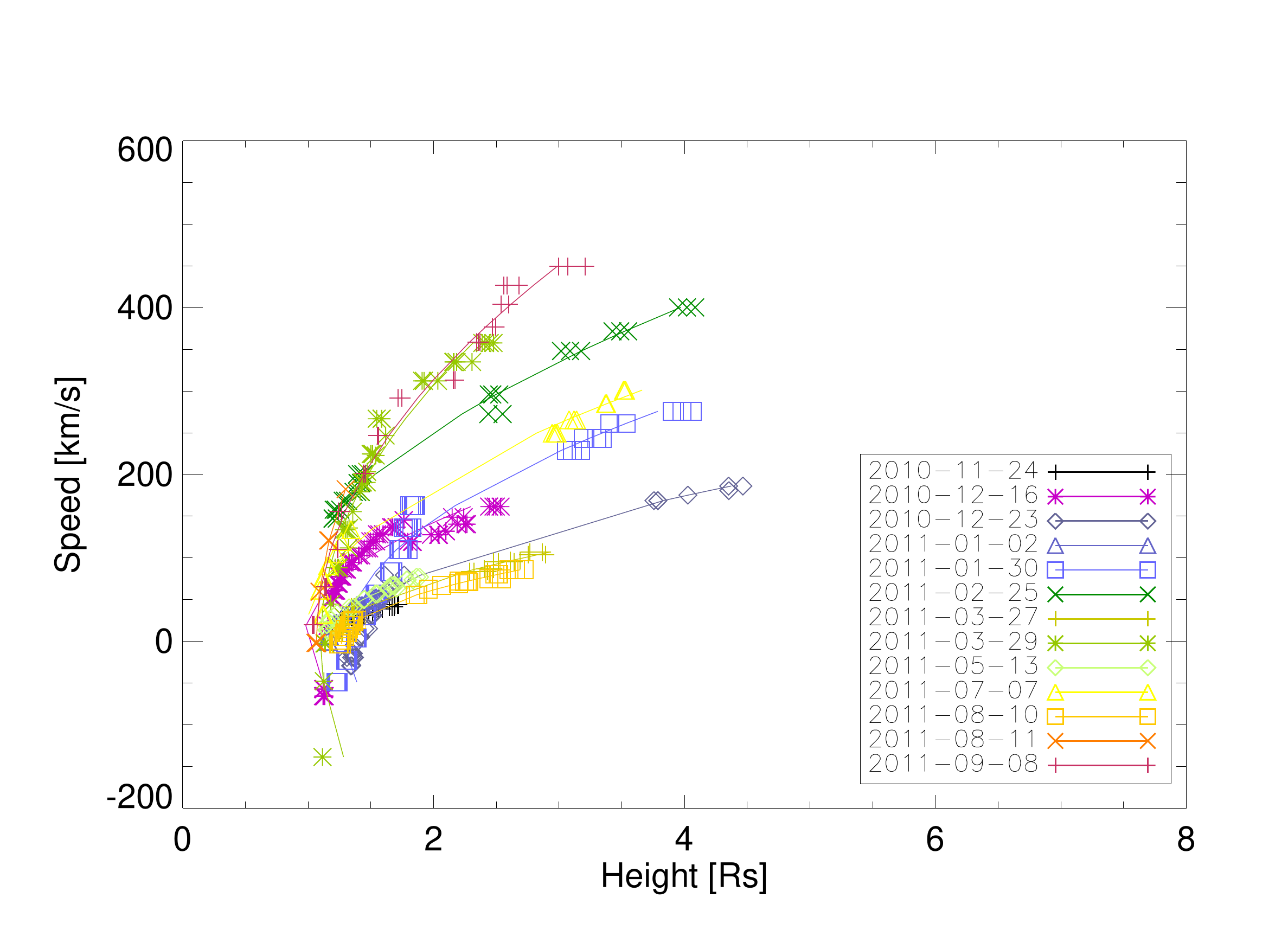}
    \hspace{0.cm}
    \includegraphics[width=0.33\textwidth,angle=90, trim=1.0cm 1.5cm 2.0cm 0.9cm, clip=true]{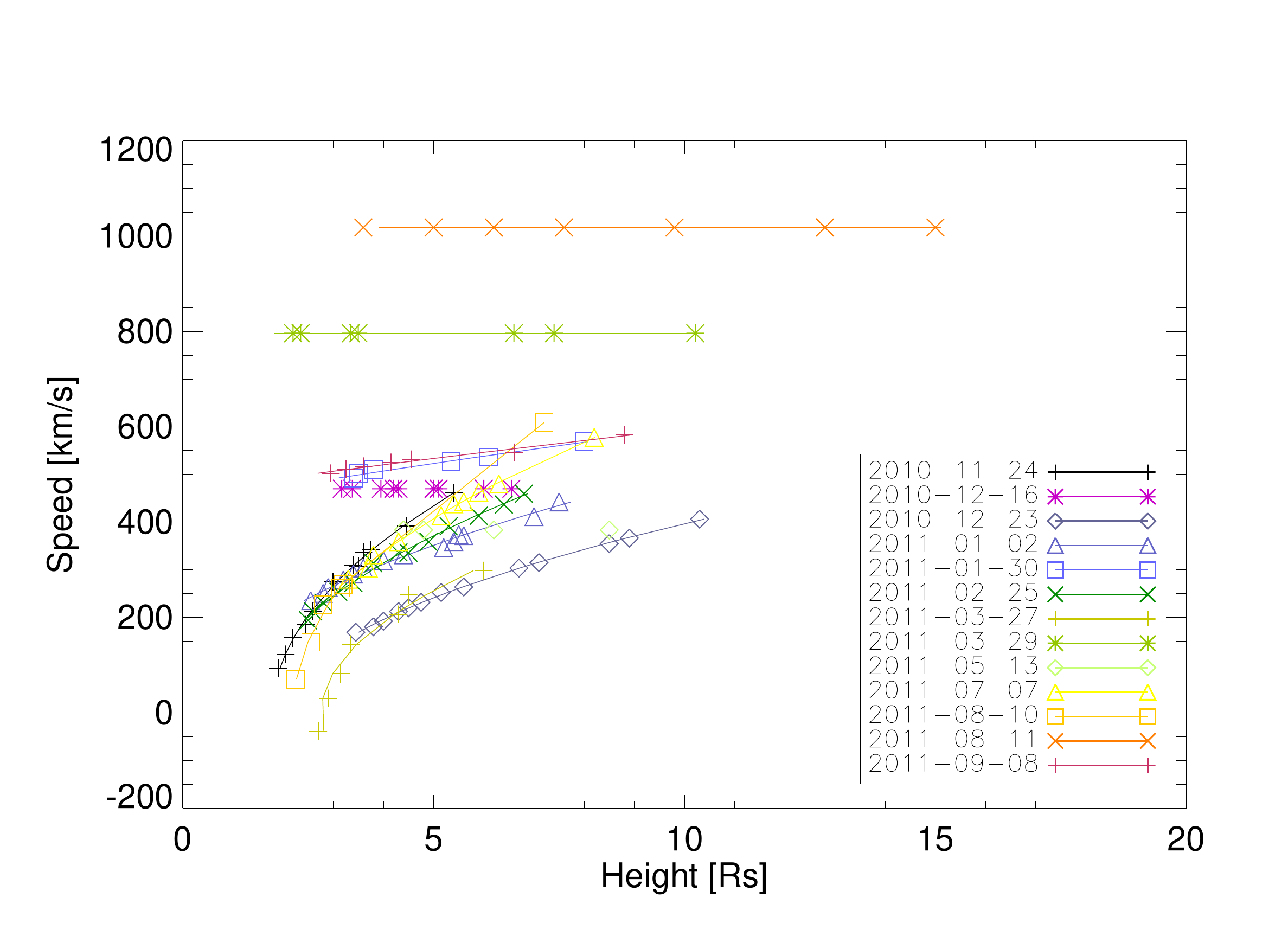}
    \caption{Radial propagation speed of prominences (left) and CMEs (right) as a function of height. The different colors indicate different events.}
    \label{fig:speeds_vs_h}
\end{figure}

With the aim of studying the relationship between propagation speed and deflection, we determined the radial speed of prominences and CMEs for all events. By applying the tie-pointing method to the apex of the prominence material and by fitting CMEs with the GCS model, both at different times, we obtained 3D coordinates as described in Section \ref{ss:meth}. We determined the radial propagation speed of prominences and CMEs by implementing quadratic or linear fits to height vs. time data. Figure~\ref{fig:speeds_vs_h} shows the resulting speeds as function of height. All prominences exhibit accelerated profiles that reach speed values of 500 km\,s$^{-1}$ while CMEs reach about 1000 km\,s$^{-1}$ with some of the events showing no acceleration. 
%
In addition, we have computed the overall 3D deflection $\Psi$ with respect to the solar source at different heights for both, prominences and CMEs. We fit an exponential function of the form $p_0-p_1 \exp(-p_2x)$ to the deflection as function of height $\Psi(h)$ using a different set of parameters for prominences and CMEs, given that in general the deflection profiles of both structures differ, and for each event. This function describes well the general behavior of the measurements, i.e. a fast increase at lower heights and a flatter trend at higher ones. There are some few CMEs that show no acceleration (right panel in Figure~\ref{fig:speeds_vs_h}), presumably because most of it took place at lower heights. 

\begin{figure}
    \centering
    \includegraphics[width=0.33\textwidth,angle=90, trim=1.0cm 1.5cm 2.0cm 0.5cm, clip=true]{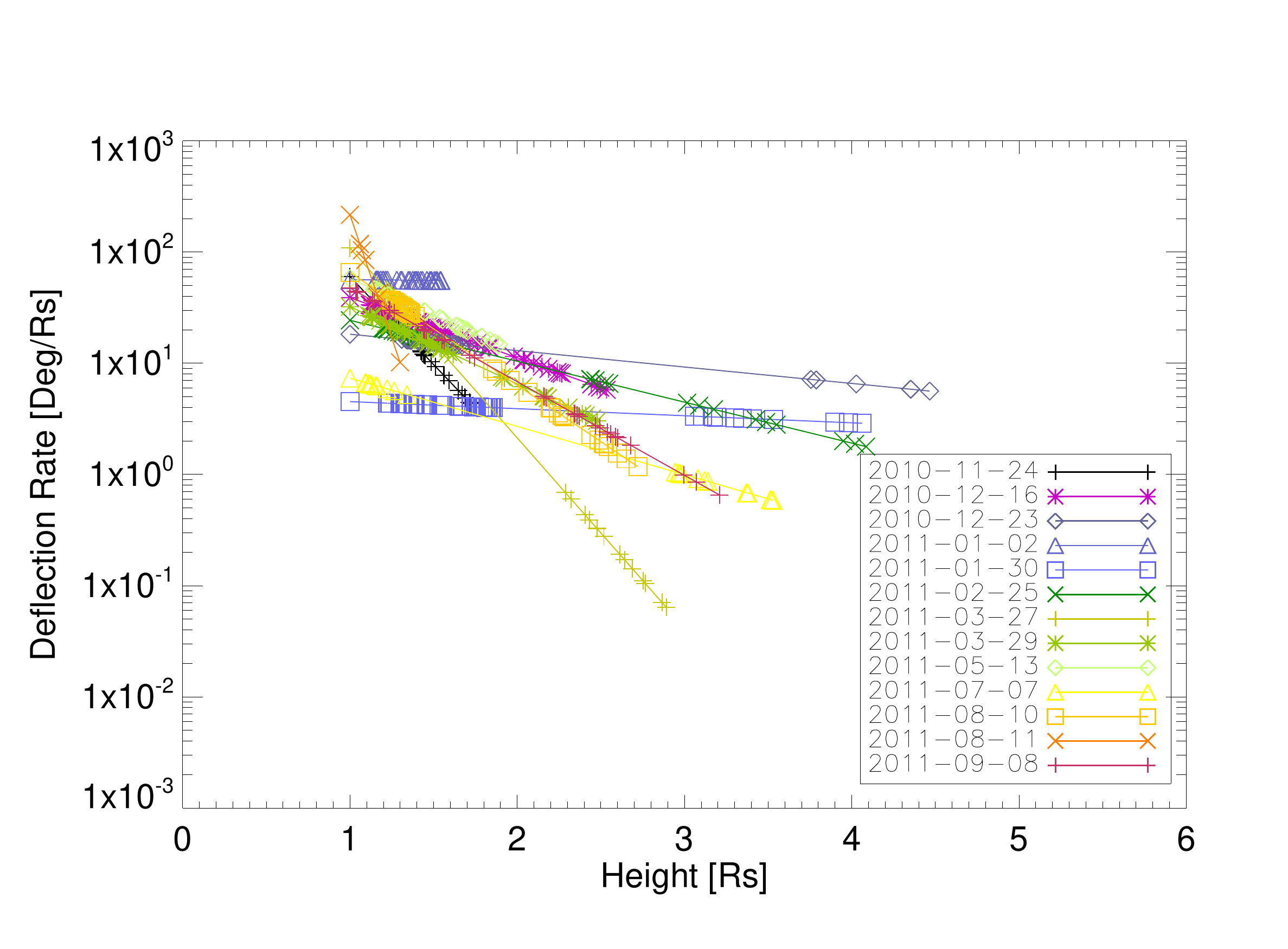}
    \includegraphics[width=0.33\textwidth,angle=90, trim=1.0cm 1.5cm 2.0cm 0.5cm, clip=true]{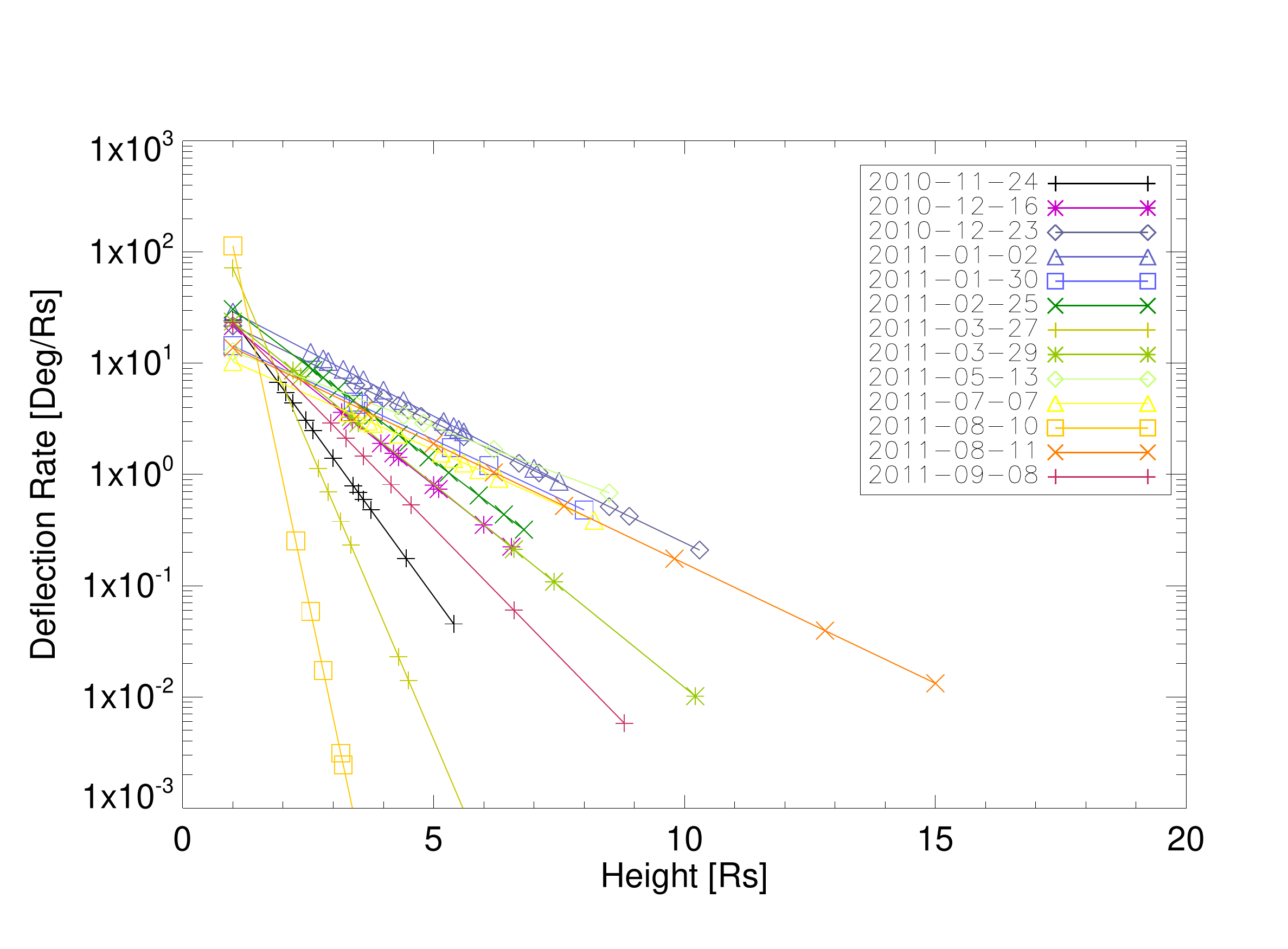}
    \caption{Deflection rate vs. height for prominences (left) and CMEs (right). The different colors indicate different events.}
    \label{fig:deflrate_vs_h}
\end{figure}

The deflection rate with height, calculated as $\frac{d \Psi}{dh}$, is shown in logarithmic scale in  Figure~\ref{fig:deflrate_vs_h}. The deflection rate for prominences (left panel) decreases abruptly, one order of magnitude  for heights lower than $2\,R_{\odot}$ for most of the events (except for events on 2011-01-02 and 2011-01-30, whose deflection rates are almost constant). In contrast, the deflection rate for CMEs (right panel of Figure~\ref{fig:deflrate_vs_h}) decreases less steeply, one order of magnitude for heights lower than  $4\,R_{\odot}$ (except for events on 2011-03-27 and 2011-08-10, which rapidly decay). Calculating the mean height where deflection rates decay $1/e$ of their initial values ($h_e$ hereafter) results in $2.3\,R_{\odot}$ for prominences and $2.4\,R_{\odot}$ for CMEs. This suggests that most of the deflection with respect to the source region occurs below $2.4\,R_{\odot}$. Also, on average, the deflection rate of prominences is greater than that of CMEs.

\begin{figure}
    \centering
    \includegraphics[width=0.33\textwidth,angle=90, trim=1.0cm 1.3cm 2.0cm 1.0cm, clip=true]{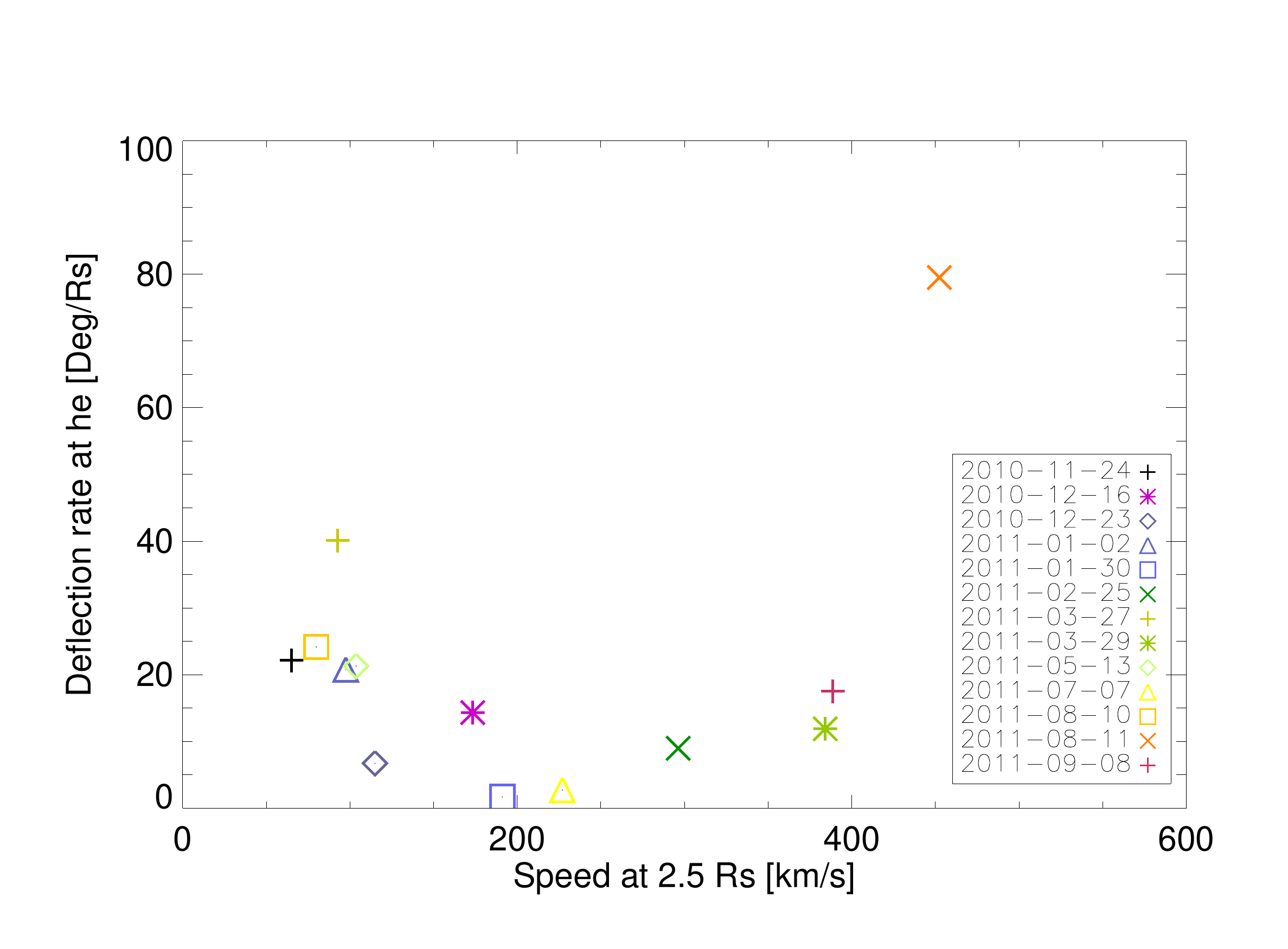}
    \hspace{0cm}
    \includegraphics[width=0.33\textwidth,angle=90, trim=1.0cm 1.5cm 2.0cm 0.5cm, clip=true]{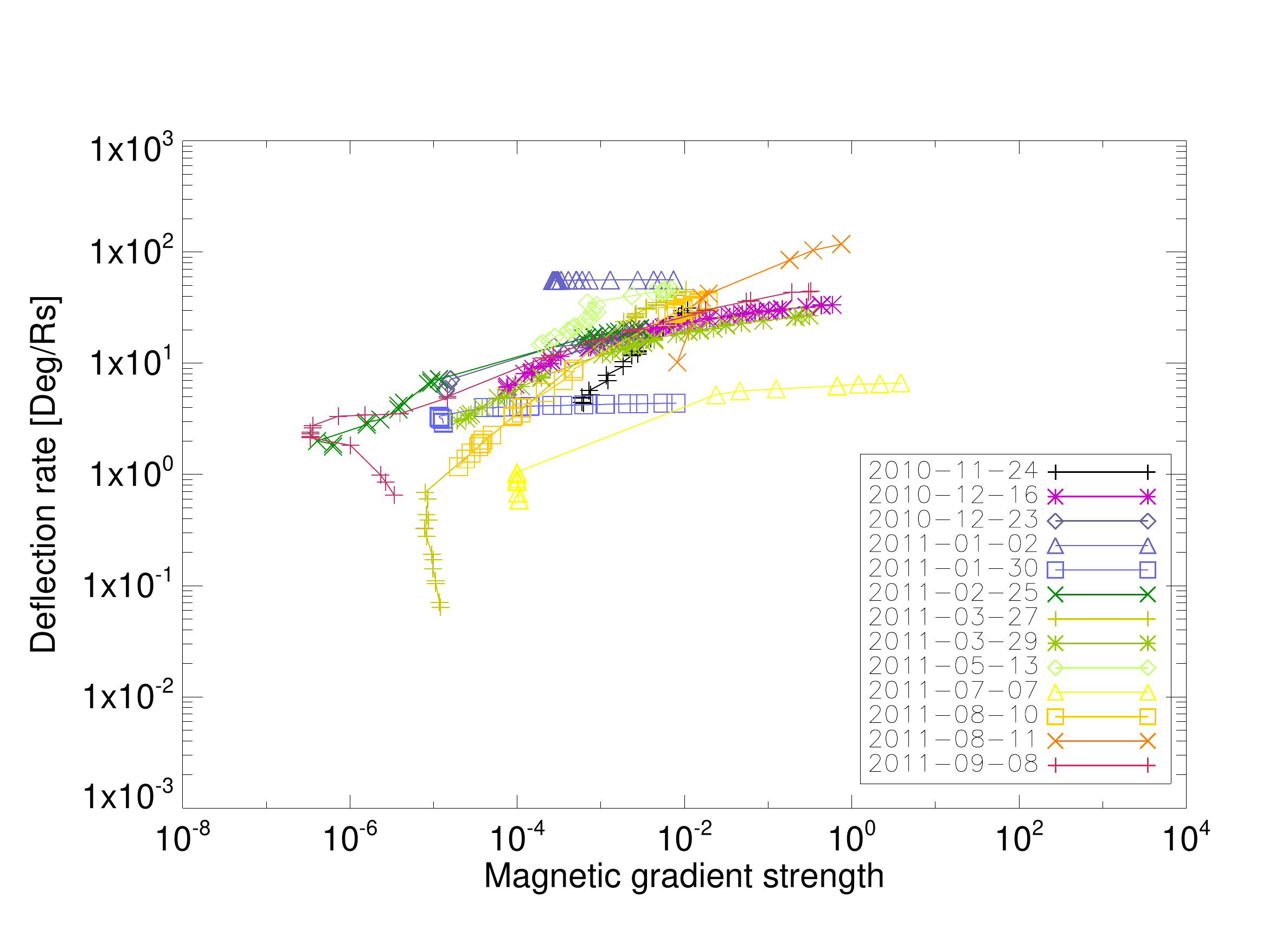}
    \caption{Left panel: Prominence deflection rate at the height where this quantity decays at $1/e$ of its initial value ($h_e$) vs. radial propagation speed at $2.5\,R_{\odot}$. Right panel: Prominence deflection rate as function of the magnitude of the magnetic gradient. The color pattern indicates different events in both panels.}
    \label{fig:deflrate_prom}
\end{figure}

To analyze in further detail the deflection rate of prominences, we show in Figure~\ref{fig:deflrate_prom} (left panel) the deflection rate at a height $h_e$ against the radial propagation speed at $2.5\,R_{\odot}$. We chose the speed value at this height because it characterizes the speed of the evolved prominence. Each event is represented by a different color. Note that in general slower prominences show deflection rates greater than 20$^\circ$, while faster events present deflection values lower than 20$^\circ$ (except 2011-08-11).   

Following the line of \citet{gui2011}, we also inspect a possible dependence between the deflection rate and the magnitude of the magnetic gradient at each latitude-longitude coordinate. Figure~\ref{fig:deflrate_prom} (right panel) displays results arising from the prominence analysis. We found a correlation factor of 0.65, suggesting that the deflection rate for prominences is proportional to the strength of the magnetic gradient. For CMEs we do not perform this analysis because we consider unchanged density maps for heights greater than $2.5\,R_{\odot}$, hence the gradient keeps its value for this height onward.    

\begin{figure}
    \centering
    \includegraphics[width=0.34\textwidth,angle=90, trim=1.0cm 1.5cm 2.0cm 1.0cm, clip=true]{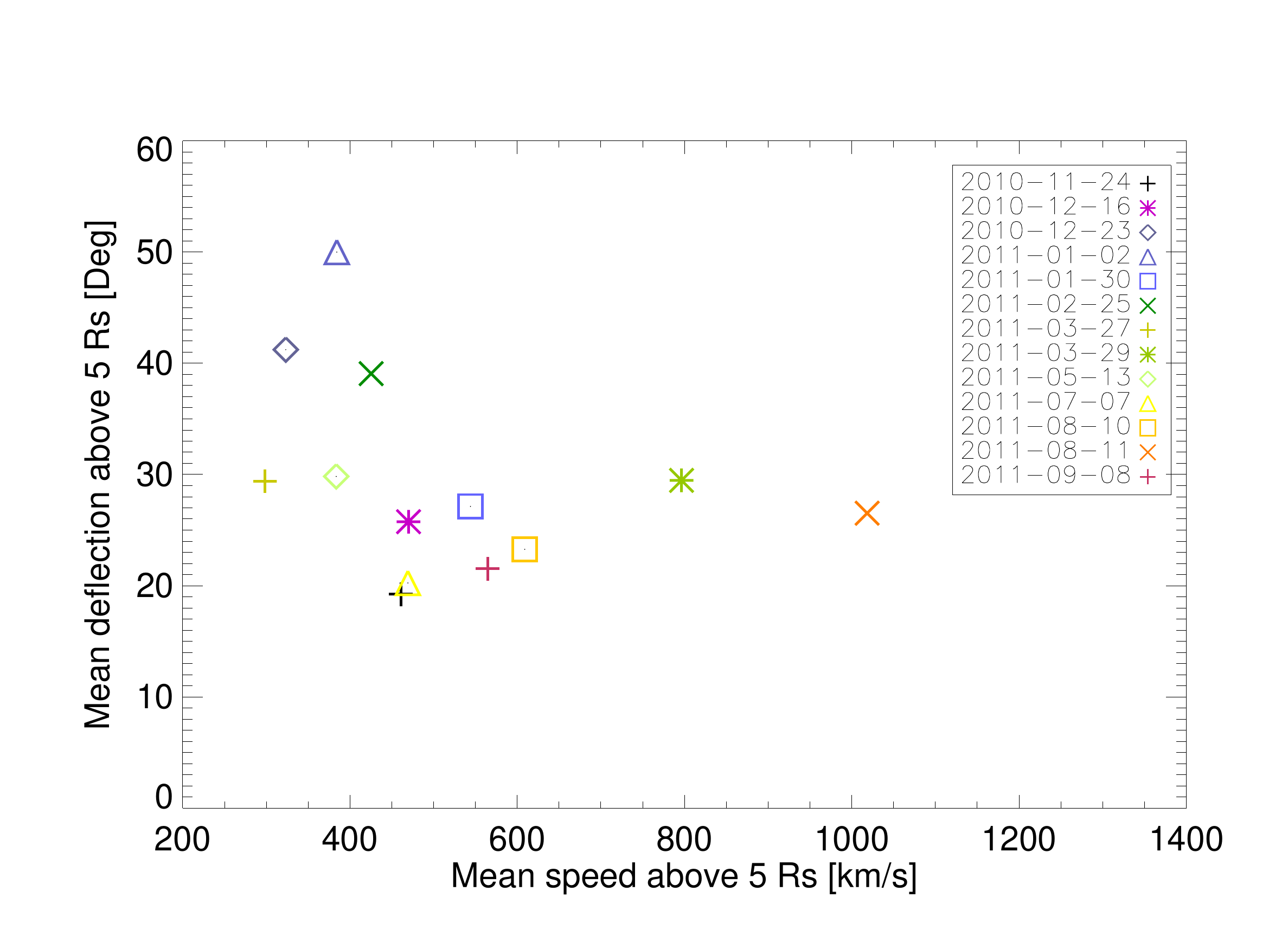}
    \caption{Total CME deflection as function of CME speed, both averaged for heights greater than $5\,R_{\odot}$.}
    \label{fig:defl_cme}
\end{figure}

For the case of CMEs, we computed the mean total 3D deflection with respect to their source regions at heights greater than $5\,R_{\odot}$, since the deflection stabilizes around that height. This overall 3D deflection is compared with the mean radial speed in Figure~\ref{fig:defl_cme}.
Two groups can be distinguished in the figure: CMEs that have speeds lower than $\sim$450 km\,s$^{-1}$ present total deflections greater than 30$^\circ$; and CMEs with speeds greater than $\sim$450 km\,s$^{-1}$ exhibit deflection values lower than 30$^\circ$.
This suggests that CMEs having speeds greater than the slow solar wind speed deflect less than slower ones.

Summarizing, the major deflection occurs at heights below $2.4\,R_{\odot}$ in the prominence domain. The deflection rate of prominences apparently is related with their radial propagation speed and the magnetic gradient strength. The total deflection with respect to the source region is presumably influenced by the CME speed relative to the solar wind speed and the direction of the magnetic field gradient.    

\subsection{Qualitative analysis of events with low $\delta$}
\label{ss:anomalous_cases}

Cases for which the angle $\delta$ (angle between CME trajectory and magnetic energy gradient) is below 120$^{\circ}$ at higher altitudes ($\sim 5\,R_{\odot}$), does not follow the general trend and thus are worth of a deeper analysis. These events are: 2011-01-30, 2011-03-27, 2011-05-13 and 2011-08-11. We can summarize qualitative findings as follows.

\subsubsection*{Events on 2011-01-30 and 2011-03-27}

\begin{figure}
    \centering
    \includegraphics[width=0.8\textwidth,angle=0]{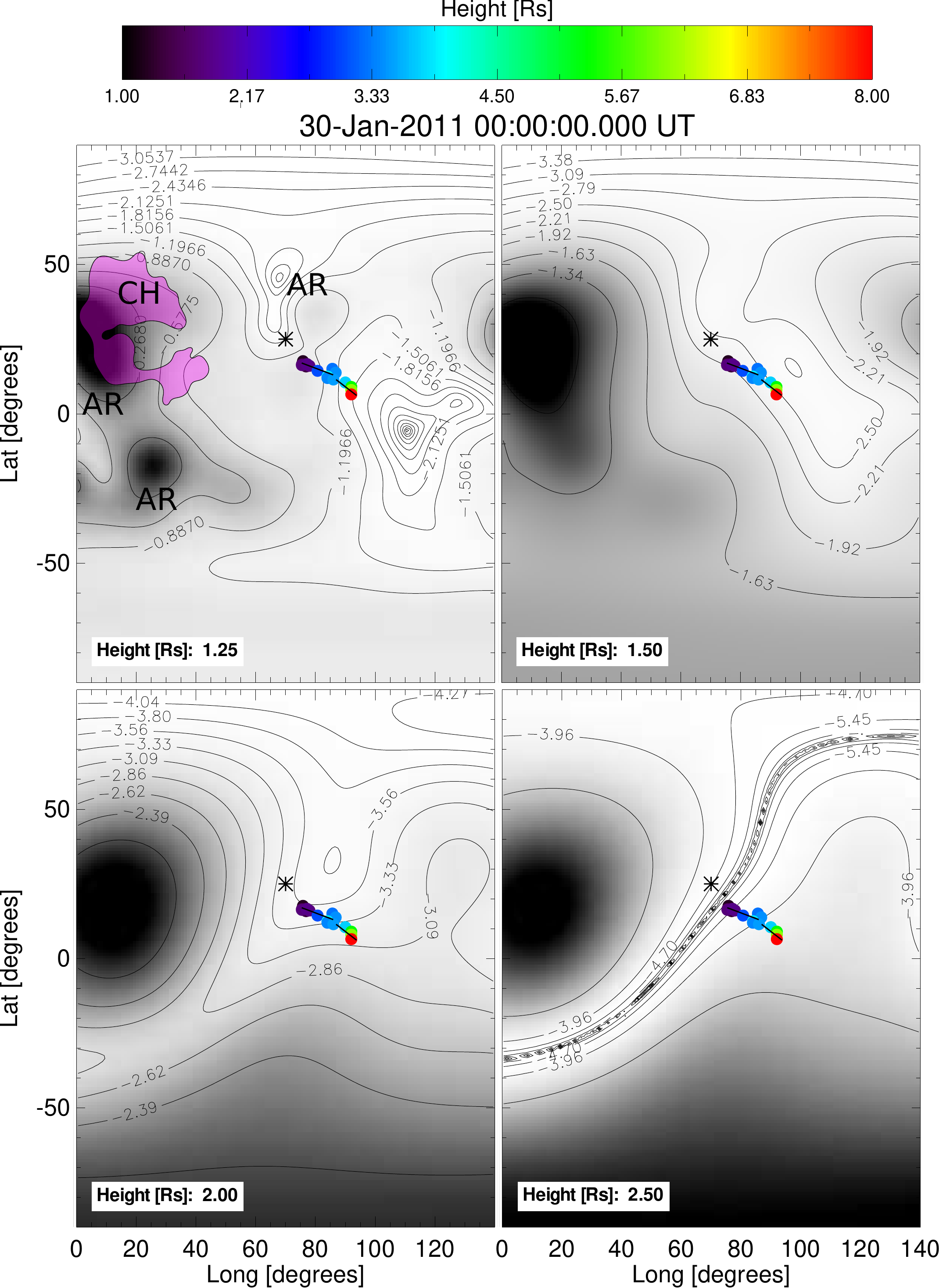}
    \caption{Magnetic  energy  density  maps  at  different  heights  for  the  event  occurred  on 30 January 2011. The corresponding height is indicated at the bottom left corner of each panel.The gray scale represents the intensity of the magnetic energy, with darker regions being those of higher intensity. Contours (solid black lines) indicate magnetic energy values in logarithmic scale. The magenta-shaded area represents the location of a CH obtained from EUV images. The  color scale of the measured points indicates their corresponding height.}
    \label{fig:2011-01-30}
\end{figure}

The CMEs on 2011-01-30 and 2011-03-27 propagate both beyond the HCS resulting in $\delta<90^{\circ}$.  Figure~\ref{fig:2011-01-30} displays the magnetic energy density maps at different heights for the event on 2011-01-30. The background gray scale represents values of magnetic energy density, where darker regions have higher strength. The colored circles indicate prominence and CME measured coordinates at various heights, while the asterisk represents the source region. Active regions and coronal holes are denoted with AR and CH, respectively. The area of the CH, obtained from EUV images, is shaded in magenta in the first map. The source region is near an AR to the north and the measured coordinates of the prominence indicate that it is first deflected toward a local magnetic energy minimum at heights lower than $1.25\,R_{\odot}$ (first panel of Figure~\ref{fig:2011-01-30}). The second and third panels ($1.5$ and $2.0\,R_{\odot}$, respectively) show that the prominence is later deflected away from the northern and eastern ARs and from the CH. Note that the CME is also moving away from these structures and follows the same initial direction. At $2.5\,R_{\odot}$ (last panel) the CME trajectory is seen beyond the HCS. On this event the influence of the magnetic energy minimum at low heights seems to be crucial for the following evolution of the structures.
\begin{figure}
    \centering
    \includegraphics[width=0.8\textwidth,angle=0]{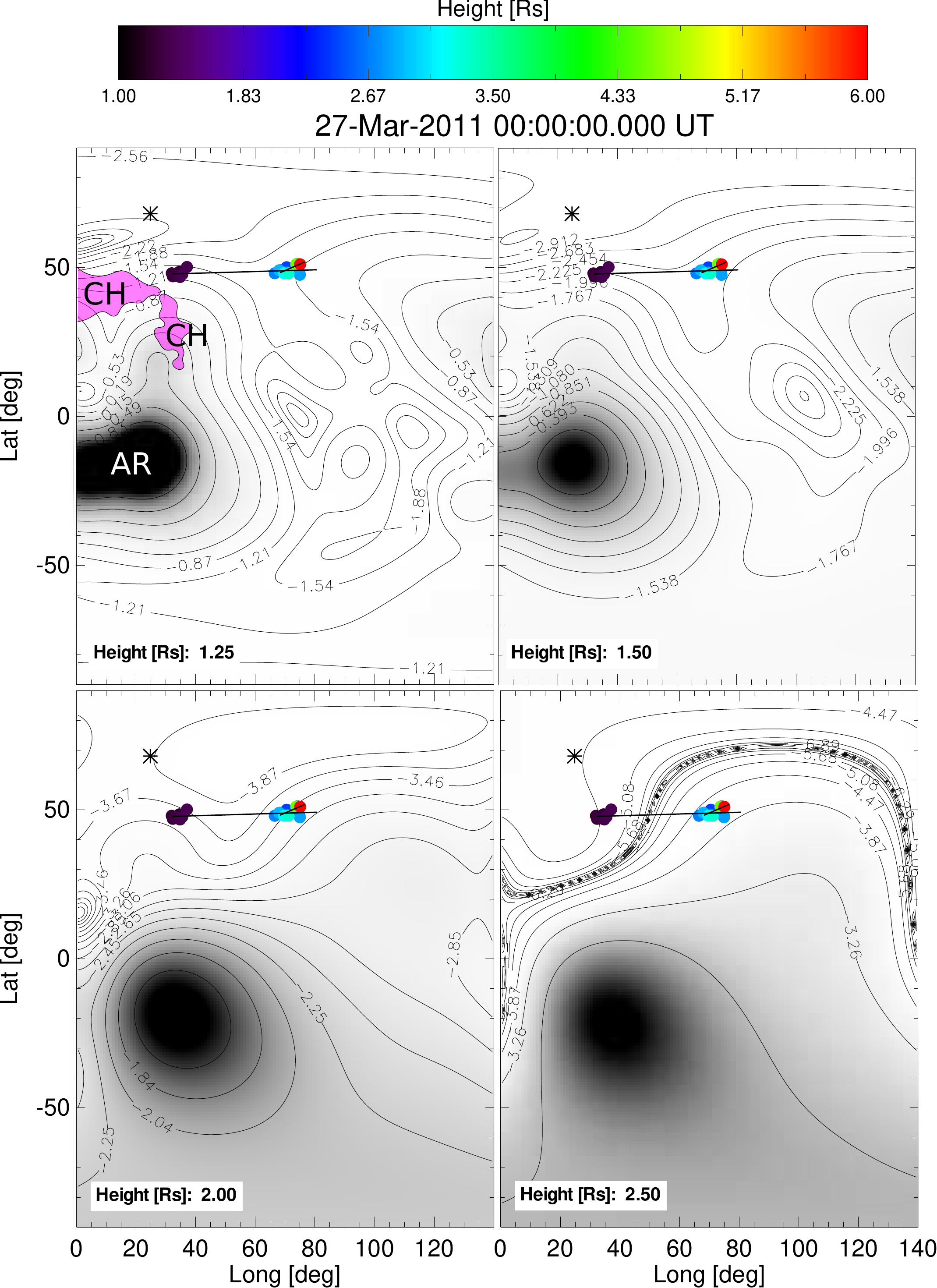}
    \caption{Idem Figure~\ref{fig:2011-01-30} but for 27 March 2011.}
    \label{fig:2011-03-27}
\end{figure}

\noindent
The initial evolution of 2011-03-27 is different. In the early stages, the prominence moves southward towards a CH (see first panel of Figure~\ref{fig:2011-03-27}) and away from open magnetic field lines located near the north pole. Then, at higher altitudes, the prominence deflects abruptly to the west moving away from the CH (second and third panel of Figure~\ref{fig:2011-03-27}), crossing the assumed location of the HCS between $1.35\,R_{\odot}$ and $2.3\,R_{\odot}$. So the CME originated beyond the HCS and its trajectory is not aligned with the magnetic energy gradient resulting in $\delta<90^{\circ}$. 
    
\noindent
As said, both of these events do not follow the path of minimum magnetic energy. In the first event the influence of the magnetic forces at low heights seems to be strong enough to push the CME beyond the HCS, in agreement with findings on some events described by \citet{kay2015}. In the second case the prominence is strongly deflected at higher altitudes by the magnetic tension of a CH. This structure does not produce a magnetic gradient variation but it would rather represent a magnetic wall that the CME is not able to penetrate, presumably because of its low speed.

\subsubsection*{The 2011-05-13 event}

\begin{figure}
    \centering
    \includegraphics[width=0.8\textwidth,angle=0]{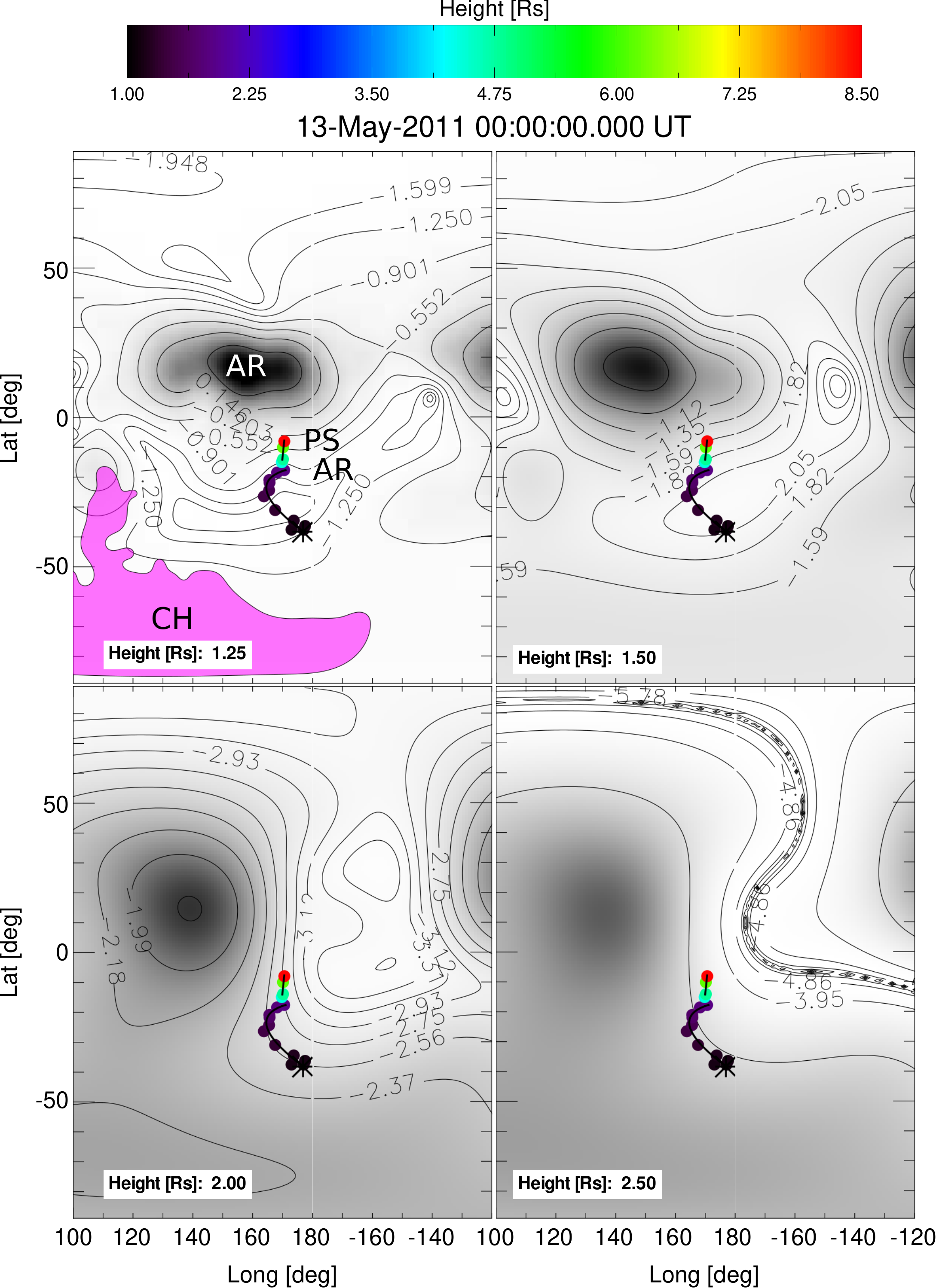}
    \caption{Idem Figure~\ref{fig:2011-01-30} but for 13 May 2011.}
    \label{fig:2011-05-13}
\end{figure}

The 2011-05-13 event shows another behavior. From the first panel of Figure~\ref{fig:2011-05-13}, we note that the prominence is located between a southern CH and a northern AR. There are also other magnetic structures surrounding the prominence: an arm of the CH located to the east between approximately $-50^{\circ}$ and $-20^{\circ}$ in latitude, and an AR and a pseudostreamer (PS) located to the west. The first and second panels of Figure~\ref{fig:2011-05-13} show that the initial trajectory of the prominence is influenced by a local minimum of magnetic energy, until $1.5R_{\odot}$, and then deflects towards lower magnetic energy region (third and fourth panel). Above $1.5R_{\odot}$, the CME moves away from the CH, presumably in an attempt to head toward regions of low magnetic energy but confined by the mentioned structures.
\begin{figure}
    \centering
    \includegraphics[width=0.8\textwidth,angle=0]{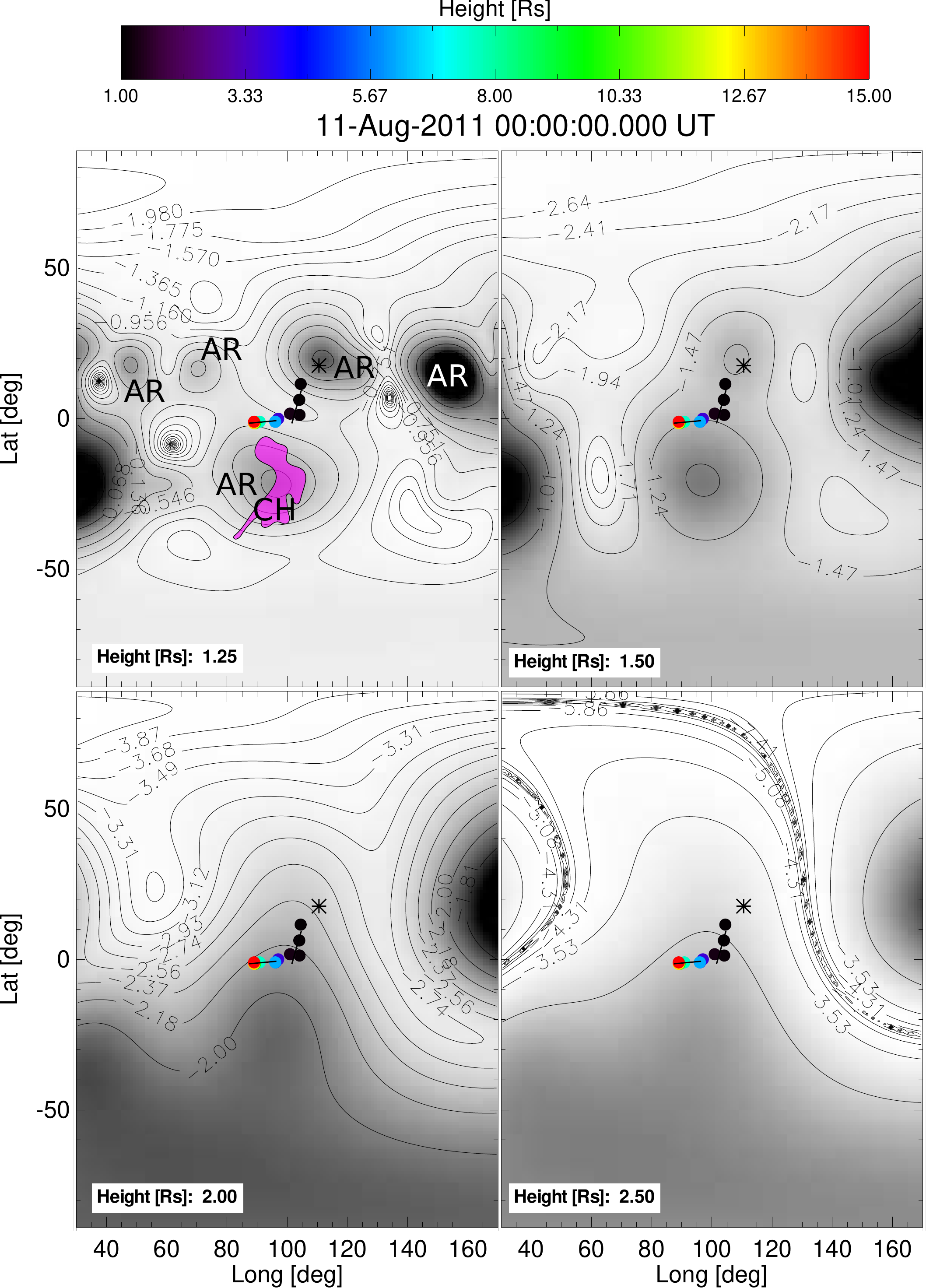}
    \caption{Idem Figure~\ref{fig:2011-01-30} but for the event on 11 August 2011.}
    \label{fig:2011-08-11}
\end{figure}

\subsubsection*{The 2011-08-11 event}

As we note from the first and second panels of Figure~\ref{fig:2011-08-11}, the source region of 2011-08-11 is an AR, and for altitudes below $1.5\,R_{\odot}$ the prominence trajectory is directed towards a local minimum of magnetic energy, moving away from the northern AR but approaching to a southern CH and an AR. This produces a deflection mainly in the latitudinal direction. At altitudes greater than $3.6\,R_{\odot}$ the CME abruptly moves towards the eastern HCS (fourth panel at $2.5\,R_{\odot}$) but not in the direction of maximum decrease of magnetic energy. This happens probably due to its high kinetic energy, given that its velocity is $1160$ km\,s$^{-1}$, which adds to the magnetic tension produced by the southern CH.

\section{Discussion and conclusions}
\label{s:discussion}

We have performed a systematic analysis of large CME deflections within a period of a year (October 2010 -- September 2011) in the rising phase of solar cycle 24. We found 13 events that deflect more than $20^{\circ}$ from their source regions. Inspired by previous reports \citep[e.g.,][]{gui2011,liewer2015,kay2015} we carried out a detailed investigation on the allegedly principal causes of deflection: the influence of magnetic forces and kinematic features. We examined these aspects from the beginning of the eruptions, studying the evolution of CMEs and their associated prominences.

To shed light on the role of these aspects, we have defined an angle $\delta$ that represents the angular span between the orientation of the trajectory of both structures and the direction of magnetic energy gradient. For prominences this angle shows disperse behavior, with half of the values aligned with the direction of minimum magnetic energy decay. Nonetheless, the deflection rate of prominences appears to be proportionally related with the magnetic gradient strength, since higher the gradient, the larger the deflection rate. This could be attributed to the fact that the magnetic structure at lower heights is more complex, with high field intensity and no large-scale structures present to affect the prominence trajectory. Other possible reasons are that the intrinsic magnetic field of the prominence and flux rope would be more intense than the surrounding magnetic structures, and also reconnection topologies and processes that are beyond the scope of this study. As a consequence of stronger magnetic fields at low altitudes, the deflection rates are larger for prominences than for CMEs, also supported by the correlation found between  deflection rate and magnetic gradient strength. The kinematic analysis revealed a tendency for slower events to have larger deflection rates (namely $>20^{\circ}/R_{\odot}$).

For CMEs we found that $\sim$70\,\% of $\delta$ values correspond to trajectories that follow directions opposite to the magnetic gradient, i.e. most of the CMEs propagate towards the minimum energy density escaping the low corona near the HCS or a region of low magnetic energy. The remaining 30\% of $\delta$ values are related to CMEs that do not obey this behavior and are analyzed in detailed in Section \ref{ss:anomalous_cases}. Possible reasons for these events not following the direction of decrease of the magnetic energy can be summarized as: a) if the source region is located close to the HCS and the magnetic forces are large at lower heights, the CME may not necessarily head toward low magnetic energy regions; b) if the CME is aimed at a region of open field lines (CH), it is abruptly deflected by the magnetic tension of this structure regardless the local magnetic pressure of the environment. In summary, we find crucial for these events the magnetic forces acting below 2.5\,R$_{\odot}$ and the magnetic tension produced by the CHs, which is not represented in the magnetic density energy maps. An additional reason for the discrepancy in the expected behavior of these events may rely on inaccuracies in the deduced locations of the HCS, currently determined from PFSS extrapolations and assuming the magnetic field is radial above 2.5\,$R_{\odot}$.

An apparently important factor related to the amount of total deflection is the speed of the CME relative to the solar wind speed. If the CME is slower than the slow solar wind speed, the total 3D deflection would be greater than for faster CMEs. This is in agreement with previous reports \citep{gui2011, kay2015}.  

The analysis performed here shows deflection both in latitude and longitude and the events exhibit a variety of behaviors, making systematization a difficult task. Studies like this for events having large deflections are necessary for other events with different characteristics and in different phases of the solar cycle. In this way, a broader view of the conditions determining that CMEs deflect or follow a radial trajectory, will be achieved. However, tracking prominences and CMEs over several moments of time and in 3D space is a difficult and time-consuming task, which is also affected by the different characteristics and limitations of the instruments used to observe structures at diverse heights.  The PROBA2/SWAP instrument concept of an extended FOV to bridge the gap between other low coronal imagers and coronagraphs is useful in this respect, as it promises to be its successor on board PROBA3 \citep{Lamy-etal2010}. In addition, coronagraphs aboard off-the-ecliptic missions, like Solar Orbiter's METIS \citep{Antonucci-etal2017}, will enable better constraining of longitudinal deflections and 3D coordinates of structures overall.

\section*{Acknowledgments}
MVS and MC acknowledge the PROBA2 Guest Investigator program grant received  to  carry  out  this  work and the SWAP data provided by the PROBA2 team. MVS acknowledges support from CONICET as postdoc fellow. MC, HC, FAI and AC are members of the Carrera del Investigador Cient\'ifico (CONICET). AS is doctoral fellow of CONICET. MC and AS acknowledge support from ANPCyT under grant number PICT No. 2016-2480. MC, MVS and AS also acknowledge support by SECYT-UNC grant number PC No 33620180101147CB. MVS, HC, and FAI thank the support of project UTN UTI4915TC. MM, MW and ED acknowledge support from the Belgian Federal Science Policy Office (BELSPO) through the ESA-PRODEX programme, grant No. 4000120800. GS acknowledges the support from the NASA STEREO/SECCHI program NNG17PP27I. 
The authors also acknowledge the use of the data from NSO/GONG, SDO/AIA, SOHO/LASCO, STEREO/EUVI, STEREO/COR1, STEREO/COR2. 

%

\bibliographystyle{plainnat}

\bibliography{bibliography}

\end{document}